\documentclass[aps,prl,amsmath,amssymb,floatfix,twocolumn,amsmath,superscriptaddress,
twocolumn,nofootinbib,tighten,letterpaper]{revtex4-2}
\usepackage{lmodern}
\usepackage[T1]{fontenc}
\usepackage[latin9]{inputenc}
\usepackage{float}
\usepackage[normalem]{ulem}
\usepackage{amsmath, dsfont}
\usepackage{amssymb}
\usepackage{mathtools}
\usepackage{mathdots}

\usepackage{graphicx}
\usepackage{wasysym}
\usepackage{color}
\usepackage[table]{xcolor}
\usepackage{bm}
\usepackage{hyperref}
\hypersetup{pdfpagemode=UseNone}
\usepackage[many]{tcolorbox}
\newsavebox\MBox

\allowdisplaybreaks

\newcommand{\bsub}{\begin{subequations}}
\newcommand{\esub}{\end{subequations}}

 \graphicspath{{Images/}}

\begin{document}

\title{Metal-insulator transition in a 2D system of chiral unitary class}

\author{Jonas F.~Karcher}
\affiliation{{Pennsylvania State University, Department of Physics, University Park, Pennsylvania 16802, USA}}
\affiliation{{Institute for Quantum Materials and Technologies, Karlsruhe Institute of Technology, 76021 Karlsruhe, Germany}}
\affiliation{{Institut f\"ur Theorie der Kondensierten Materie, Karlsruhe Institute of Technology, 76128 Karlsruhe, Germany}}

\author{Ilya A.~Gruzberg}
\affiliation{Ohio State University, Department of Physics, 191 West Woodruff Ave, Columbus OH, 43210, USA}

\author{Alexander D.~Mirlin}
\affiliation{{Institute for Quantum Materials and Technologies, Karlsruhe Institute of Technology, 76021 Karlsruhe, Germany}}
\affiliation{{Institut f\"ur Theorie der Kondensierten Materie, Karlsruhe Institute of Technology, 76128 Karlsruhe, Germany}}

\date{September 23, 2022}

\begin{abstract}
We perform a numerical investigation of Anderson metal-insulator transition (MIT)  in a two-dimensional system of chiral symmetry class AIII by combining finite-size scaling, transport, density of states, and multifractality studies. The results are in agreement with the sigma-model renormalization-group theory, where MIT is driven by proliferation of vortices.  We determine the phase diagram and find an apparent non-universality of several parameters on the critical line of MIT, which is consistent with the analytically predicted slow renormalization towards the ultimate fixed point of the MIT. The localization-length exponent $\nu$ is estimated as $\nu = 1.55 \pm 0.1$.
\end{abstract}

\maketitle

{\it Introduction.}  Anderson transitions {(ATs)} in disordered systems---which include metal-insulator transitions (MITs) as well as transitions between topologically distinct insulating phases---remain a dynamic field of research~\cite{evers08}. In this context, two-dimensional (2D) systems attract particular attention. On the experimental side, there is a variety of realizations of 2D electronic disordered systems, including semiconductor heterostructures and MOSFETs, graphene and other 2D materials, oxide heterostructures, as well as surfaces of topological insulators and superconductors.  Furthermore, investigation of 2D disordered systems in photonic structures is an emerging research area~\cite{stuetzer2018photonic}.

For the most conventional setting of a quantum particle in a random potential (Wigner-Dyson orthogonal symmetry class AI),  $d=2$ is a lower critical dimensionality, as for conventional second-order phase transitions with continuous symmetry. This implies that there is no {AT} in 2D systems of this symmetry class and all states are localized (although the localization length is exponentially large for weak disorder). At the same time, it {was realized}
that there is a number of mechanisms generating {ATs} in 2D disordered systems of other symmetry classes.  While field theories of {ATs} are non-linear sigma models with a continuous non-abelian symmetry, the existence of metallic (symmetry-broken) phases in 2D geometry is not in conflict with the Mermin-Wagner theorem, in view of an unconventional character of the symmetry groups (involving supersymmetry and non-compactness or replica limit, depending on the formulation).

The 2D {ATs} include, in particular, MITs in classes AII, D, and DIII with broken spin-rotation invariance playing a crucial role, as well as quantum-Hall transitions in classes A, C, and D that are governed by topology.  Whereas {ATs} of these types have been studied in a rather detailed fashion, there is one more type of 2D {ATs} that has received much less attention: MITs in chiral classes AIII, BDI, and CII. In fact, early studies demonstrated a resilience of chiral systems to Anderson localization, leading to a suggestion that 2D and 3D systems of chiral symmetry classes do not exhibit {AT} at all, remaining always in a delocalized phase~\cite{antoniou1977absence}. This has received an apparent support from the renormalization-group (RG) analysis of the corresponding sigma models performed in pioneering works of Gade and Wegner~\cite{gade1991the,gade1993anderson}, which yielded no quantum corrections to conductivity (and thus no localization) to {all orders in} perturbation theory. The Gade-Wegner RG implies that 2D systems of chiral classes possess a metallic phase with a line of infrared-stable fixed points with different values of conductivity.  The special character of RG in chiral classes is related to the fact that the corresponding sigma-model manifolds contain an additional U(1) degree of freedom.

More recently, numerical studies of suitably designed 2D models of chiral classes have provided evidence of Anderson MITs~\cite{motrunich2002particle-hole,bocquet2003network}. An analytical theory of 2D {ATs} in chiral classes was developed in Ref.~\cite{koenig2012metal-insulator}. It was pointed out in Ref.~\cite{koenig2012metal-insulator} that, since the sigma-model manifolds for chiral classes are not simply connected [due to {the} U(1) degree of freedom], they allow for topological excitations---vortices. Inclusion of the vortices in the RG analysis leads to a metal-insulator phase transition~\cite{koenig2012metal-insulator}, in an analogy with the famous Berezinskii-Kosterlitz-Thouless (BKT) transition in {the} XY model. The analysis of the resulting RG flow showed, however, that there is an essential difference: the transition happens at a finite fugacity $y > 0$, at variance with the fixed point value $y=0$ for the BKT transition. This hinders a fully controllable analytical calculation of critical exponents at
{MITs}
in chiral classes, thus making numerical studies of these transitions even more important. The central goal of this paper is a numerical study of the 2D MIT in the chiral unitary class AIII.

{\it Chiral classes.}  The special character of disordered systems of chiral symmetry classes has been understood since the pioneering work of Dyson who found a singularity of the density of states in 1D harmonic chains at zero energy (chiral symmetry point)~\cite{dyson1953dynamics}.  Further works extended the analysis to localization properties and to quasi-1D systems. It was found that an $N$-channel quasi-1D system of chiral class  has $N$ topological phases. At transitions between these phases,  the density of states exhibits the Dyson singularity~\cite{mckenzie1996exact, titov2001fokker}, and the localization length diverges~\cite{theodorou1976extended, eggarter1978singular, fisher1995critical, balents1997delocalization, brouwer1998delocalization, mudry1999random, brouwer2000nonuniversality, altland2001spectral, altland2015topology}.  Critical points of these transitions have infinite-randomness character, with critical wave functions showing very strong fluctuations~\cite{balents1997delocalization, karcher2019disorder}.

For 2D chiral-class systems, most of the past research focussed on properties of the metallic phase. The Gade-Wegner sigma model was re-derived and analyzed in many works~\cite{altland1999field, fabrizio2000anderson, fukui1999critical, guruswami2000super}.
A particular attention was paid to the asymptotic infrared behavior, with is of infinite-randomness character, exhibiting a very strong divergence of the density of states and a ``freezing'' of the multifractality spectrum~\cite{motrunich2002particle-hole, mudry2003density, yamada2004random, dellanna2006anomalous}. On {the} numerical side, most papers showed critical properties of the metallic phase that are characterized by non-universal exponents for various observables  (such as multifractality, density of states, localization length at finite energy)~\cite{hatsugai1997disordered, cerovski2000bond-disordered, eilmes2001exponents, eilmes2004exponents, markos2007critical, schweitzer2008disorder-driven} and are essentially different from those expected in the infinite-randomness infrared limit. This is not surprising: the Gade-Wegner flow towards the line of infrared fixed points is logarithmically slow, so that in a typical situation the infrared limiting behavior can likely be out of reach on any realistic length scale. In several works~\cite{markos2010logarithmic, schweitzer2012scaling, markos2012disordered}, evidence of the asymptotic behavior of the lowest Lyapunov exponents in the quasi-1D geometry has been reported.

Apart from realizations in disordered electronic systems, the interest to models {in the} chiral classes is due to their relation to models of Dirac fermions coupled to fluctuating gauge fileds that are discussed in the context of quantum chromodynamics (QCD)~\cite{verbaarschot1994spectrum}. It was proposed that {ATs} in such models may be connected to QCD phase transitions; see Ref.~\cite{giordano2021localization} for a recent review. It is also worth mentioning that chiral-class models can be experimentally realized
in microwave setups based on coupled resonators~\cite{rehemanjiang2020microwave}. Recently, MITs in 3D chiral-class systems were studied in Refs.~\cite{luo2020critical,wang2021universality}.

{\it Field theory of 2D chiral AT.}   In the fermionic replica formalism, the sigma-model manifolds for classes AIII, BDI, and CII are $\text{U}(n)$, $\text{U}(2n)/\text{Sp}(2n)$, and $\text{U}(n)/\text{O}(n)$, respectively. In the analytical and numerical analysis below, we focus on the class AIII. The Gade-Wegner sigma-model action has the form~\cite{gade1991the,gade1993anderson}
$$
S[Q] = - \! \int d^2 r \left[ \frac{\sigma}{8\pi} \text{Tr} (U^{-1} \nabla U)^2 + \frac{\kappa}{8\pi} (\text{Tr} U^{-1} \nabla U)^2 \right].
$$
Here $U(\mathbf{r}) \in \text{U}(n)$ (with the replica limit $n \to 0$ to be taken at the end of the calculation), $\sigma$ is the conductivity in units of $e^2/ \pi h$; the second term (known as ``{the} Gade term'') couples only to the U(1) degree of freedom and is specific for chiral classes. To describe the transition, one has to include also vortices, with a fugacity $y$~\cite{koenig2012metal-insulator}. The RG equations for three couplings $\sigma$, $\kappa$, and $y$ read
\begin{align}
\partial K / \partial \ln L &= 1/4 -2Ky^2  \,,
\label{eq:RG-K} \\
\partial y / \partial \ln L &= (2-K)y \,, \label{eq:RG-y}  \\
\partial \sigma / \partial \ln L &= -\sigma y^2 \,,  \label{eq:RG-sigma}
\end{align}
where $K= (\sigma + \kappa)/4$.  Equations~\eqref{eq:RG-K}  and {\eqref{eq:RG-y}} form a closed system, with a fixed point at $K=2$ and $y=\frac14$. In the three-dimensional parameter space ($\sigma$, $\kappa$, $y$),  this corresponds to a critical line of {MITs}, $\sigma + \kappa = 8$, $y=\frac14$. Along this line, there is a flow according to Eq.~\eqref{eq:RG-sigma} towards the ultimate fixed point $\sigma =0$, $\kappa = 8$, and $y=\frac14$.  This flow is, however, very slow:  $\sigma(L) {=} \sigma_0 L^{-1/16}$.  Therefore, while in the strict infrared limit the transition is described by the ultimate fixed point, on realistic scales one expects to see a transition described by some point on the critical line. This is expected to lead to an apparent non-universality of some of critical properties, as discussed below.

\begin{figure}
	\centering
	\includegraphics[width =.45\textwidth]{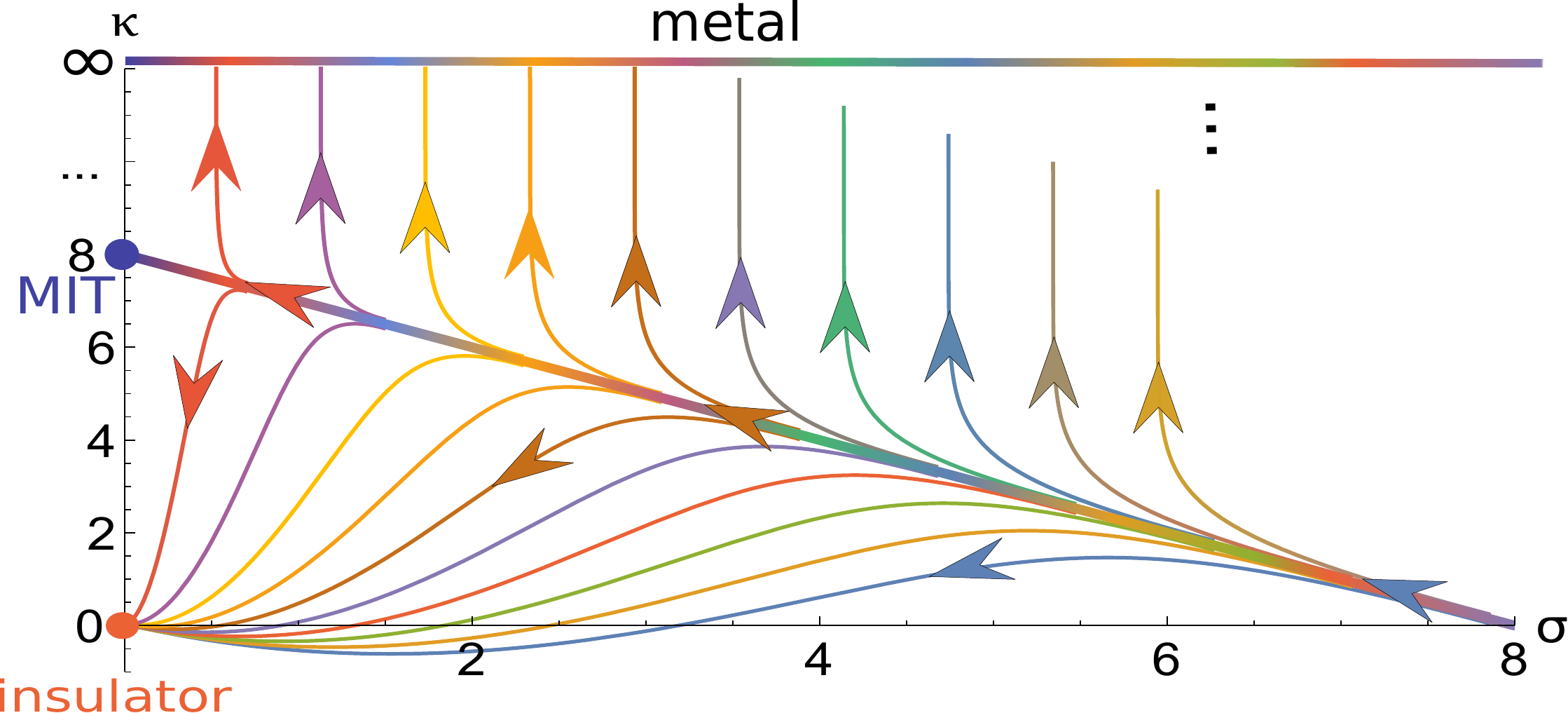}
	\caption{Schematic representation of the RG flow implied by Eqs.~\eqref{eq:RG-K}-\eqref{eq:RG-sigma}. The starting value $y_0$ of the fugacity is taken to be the critical one, $y_0 = \frac14$, and the resulting flow is projected to the $\sigma$ -- $\kappa$ plane. }
	\label{fig:RG-flow}
\end{figure}

The RG flow that follows from Eqs.~\eqref{eq:RG-K}--\eqref{eq:RG-sigma} is illustrated in Fig.~\ref{fig:RG-flow}.
The overall flow is three-dimensional and is thus difficult to display. What is shown is {the} projection of the flow on the $\sigma$--$\kappa$ plane, with all RG trajectories having an initial value of the fugacity $y_0= \frac14$. The fixed points of the flow are as follows. First, there is an infrared-stable line of fixed points describing the metallic phase, with $\sigma$ being an arbitrary constant, $\kappa \to \infty$, $y \to 0$.  Second, there is an infrared-stable fixed point describing the insulating phase: $\sigma, \kappa \to 0$ and $y \to \infty$. Finally, there is a fixed point $\sigma=0$, $\kappa=8$, $y= \frac14$, describing the MIT. It has one unstable direction, so that there is a two-dimensional critical surface with a flow towards this point. A cross-section of this surface with the plane $y=\frac14$ is the critical line $\sigma + \kappa = 8$ shown in Fig.~\ref{fig:RG-flow}.

Linearizing the RG equations~\eqref{eq:RG-K} and~\eqref{eq:RG-y} near the transition point, we get the critical exponent of the localization length $\nu =1.54$ and the irrelevant exponent $y_\text{irr}=0.77$.  In addition, there is a very slow flow towards the fixed point along the critical line described by Eq.~\eqref{eq:RG-sigma}; it yields an exponent { $y'_\text{irr} = 1/16 \simeq 0.06$}.
The fact that the ultimate fixed point of the transition is at $\sigma=0$ implies very strong fluctuations of critical eigenfunctions in the infrared limit (with freezing of the multifractal spectrum). This is expected on physical grounds: we know that eigenstates in the metallic phase possess this property, and it would be surprising if eigenstates at the transition would be ``less localized'' than in the metal.

Let us reiterate that the RG equations are only controllable at $y \ll 1$.  Since the fixed point of the transition is {at $y= \frac14$}, that is not parametrically small, all quantitative conclusions about the transition should be taken with caution. A plausible assumption is that the obtained flow is qualitatively correct but numbers describing the transition may differ substantially. It is thus crucially important to explore the transition numerically, which is done below.

{\it Model.}
We study the bipartite Hamiltonian defined on a square lattice,
\begin{align}
H &= \sum_{i,j}  \left[ c_{i,j}^\dagger t^{(x)}_{i,j} c_{i+1,j} + c_{i.j}^\dagger t^{(y)}_{i,j} c_{i,j+1} + \text{h.c.} \right],
\label{eq:ham_aiii}
\end{align}
with hoppings
{
\begin{align}
t^{(x)}_{i,j} &= \big(1 + \tfrac12 (e^{-\delta} - 1)[(-1)^i + 1] \big)({1} + v_{i,j}),
\nonumber \\
t^{(y)}_{i,j} &= \big(1 + \tfrac12 (e^{-\delta} - 1)[(-1)^j + 1] \big)({1} + w_{i,j}).
\end{align}
}
Disorder is introduced via random $v_{i,j}$ and $w_{i,j}$, whose real and imaginary parts are drawn independently from box distributions on $[-W/2, W/2]$. Since the matrix elements are complex, the time-reversal symmetry is broken, which puts $H$ in the chiral unitary class AIII. The real parameter $\delta$ controls the degree of staggering, which is absent for $\delta=0$ and maximal for $\delta\rightarrow \pm \infty$, when the system decouples into $2\times 2$ plaquettes.

{\it Finite-size scaling.} To locate the MIT, we use the transfer-matrix method for a quasi-1D strip of width $M= 12, \ldots, 256$ and large length $L=10^5$, with periodic boundary conditions in the transverse ($M$) direction. The extracted Lyapunov exponents $\lambda_{k,M}$ become self-averaging at large $L$. The inverse of the smallest Lyapunov exponent yields the quasi-1D localization length $\xi_M = \lambda^{-1}_{0,M}$. In the localized phase, $\xi_M$ is determined, for large $M$, by the 2D localization length $\xi_{\text{2D}}$, so that $\xi_M/M \to 0$ at $M\to \infty$. In contrast, in the metallic phase, the large-$M$ limit of $\xi_M /M$ is nonzero. { Note that this limit is finite (at variance with conventional MITs), which reflects a peculiar critical nature of the metallic phase in 2D chiral-class systems.}

\begin{figure}
	\centering
	\includegraphics[width=0.42\textwidth]{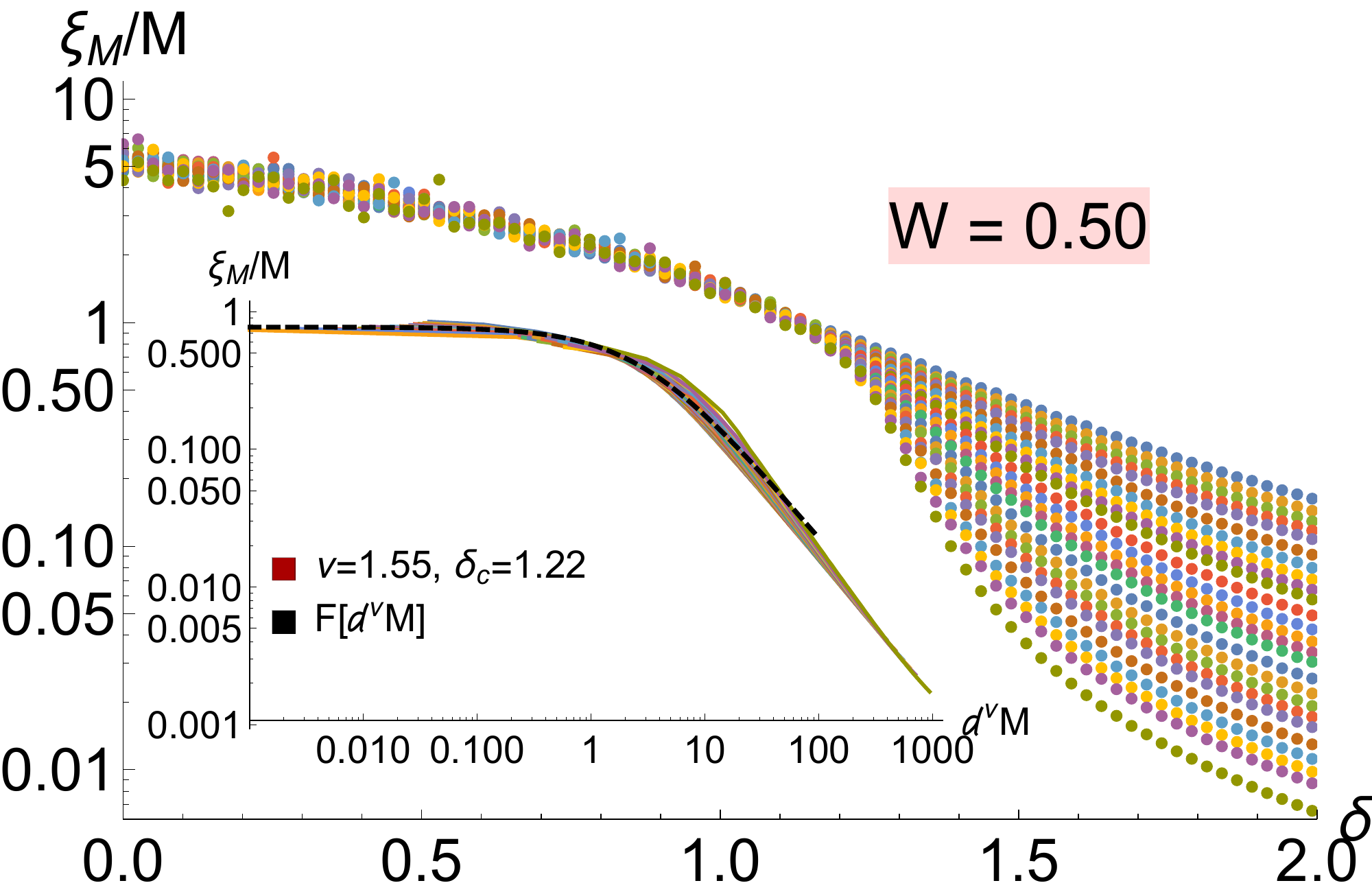}
	\caption{Finite-size scaling analysis. Ratio $\xi_M/M$ as a function of staggering $\delta$ for disorder $W=0.5$ and $M=12, \ldots,256$. Inset: data collapse $\xi_M/M = F(d^\nu M)$, with $d = \delta-\delta_c$, critical staggering $\delta_c = 1.22$, and the exponent $\nu =1.55$.
}
	\label{fig:finite-size-scaling}
\end{figure}

\begin{figure}
	\includegraphics[width=0.85\linewidth]{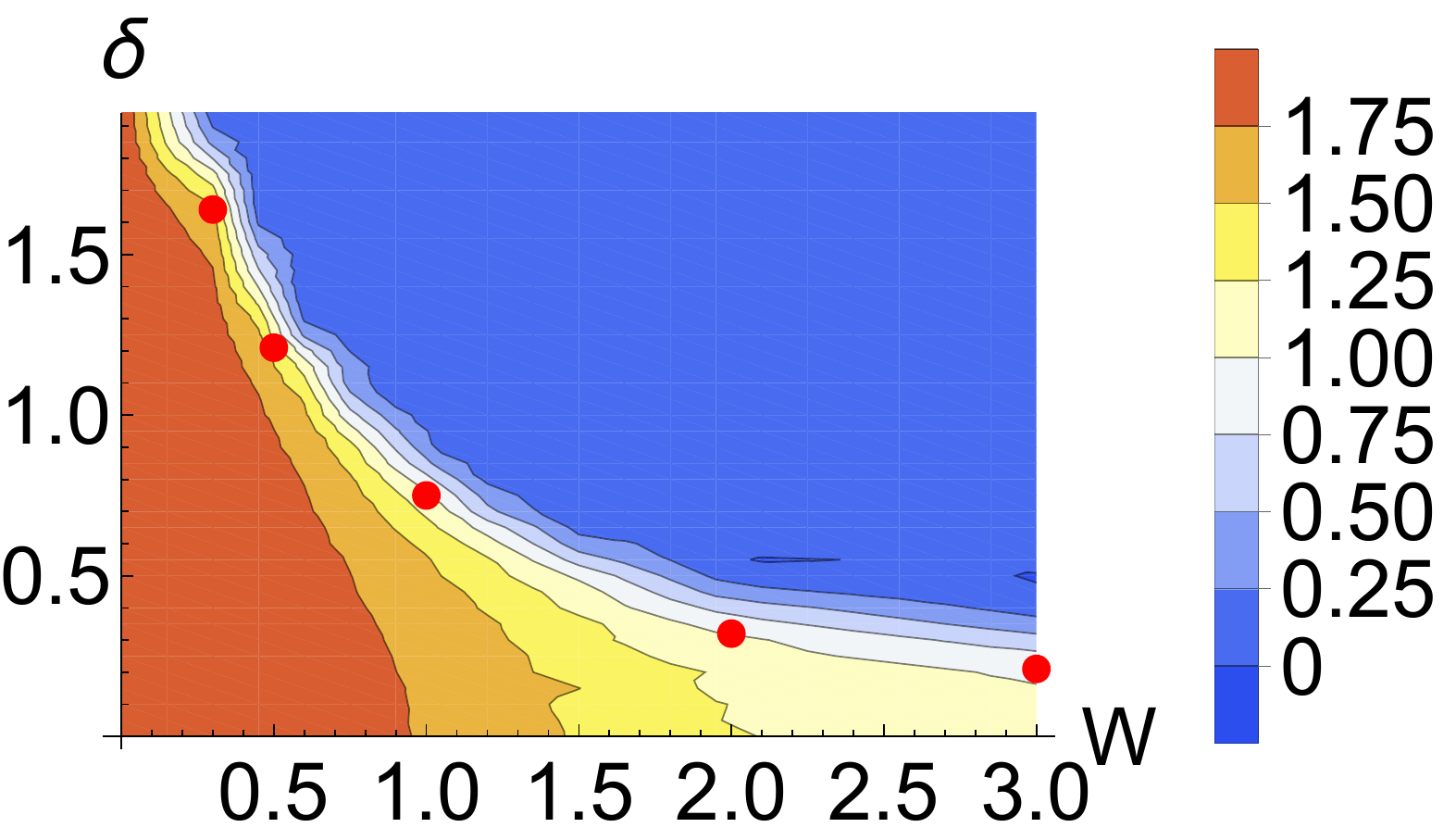}
	\caption{Phase diagram of MIT in the ($W$, $\delta$) plane.  Red symbols: MIT critical values $\delta_c(W)$ obtained by transfer-matrix analysis, see Fig.~\ref{fig:finite-size-scaling}. Color code: IPR exponent, $\tau_2(L) = - \partial \ln P_2(L) / \partial \ln L$ for largest available $L$.
	}
	\label{fig:phase}
\end{figure}

In Fig.~\ref{fig:finite-size-scaling}, we show the ratio $\xi_M/M$ for various $M$ as a function of $\delta$ for $W=0.5$. The plot clearly shows an MIT at $\delta_c \simeq 1.2$.
The same analysis is carried out for $W=0.3, 1.0, 2.0, 3.0$, see Supplementary Material (SM)~\cite{SuppMat}. The resulting phase diagram is shown in Fig.~\ref{fig:phase}.
Applying a scaling fit (see inset of Fig.~\ref{fig:finite-size-scaling}), $\xi_M/M = F(d^\nu M)$ with $d = \delta - \delta_c$, we find the exponent of the localization length, $\nu =1.55 \pm 0.1$, in a remarkable agreement with the value $\nu=1.54$ obtained from the RG equations~\eqref{eq:RG-K} and~\eqref{eq:RG-y}. A very close result for $\nu$ was obtained very recently for a related non-Hermitian model~\cite{luo2022unifying}.

Let us emphasize an apparent non-universality of the ratio $\xi_M/M$ at criticality, see Table~\ref{tab:critical-parameters}. This is consistent with a very slow RG flow along the critical line $\sigma + \kappa = 8$ predicted analytically.

\begin{table}
	\centering
\begin{tabular}{c|ccccc}
$W$ & $\delta_c$ & $\xi_M/M$& $\alpha_\nu$ & $\sigma$ & \ $ \  1/[2\pi(\alpha_0-2+x_\nu)]$\ \\
\hline
0.3 &  \ 1.64 \  & \ 0.72\  & \ $0.015$ \   & \ 3.3\  & \ 0.70\ \\
0.5 &  1.22 & 0.73 & $-0.004$ & 3.6 & 0.71\\
1.0 &  0.73 & 0.41 & $-0.025$ & 2.9 & 0.44\\
2.0 &  0.33  & 0.45 & $-0.09$ & 2.7 & 0.45\\
3.0 &  0.22 & 0.42 & $-0.11$ & 2.6 & 0.40
\end{tabular}
\caption{Critical parameters on the MIT line.
	}
	\label{tab:critical-parameters}
\end{table}

{\it Inverse participation ratio (IPR).} A complementary approach is to study directly properties of eigenstates $\psi({\bf r})$ of a 2D system. We performed {the} exact diagonalization of $L\times L$ systems with $L=24,\ldots, 768$ and periodic boundary conditions, averaging over $N=500$ disorder {realizations} and over all $L^2$ points ${\bf r} = (i,j)$ in the system. Detailed results for the averaged IPR  $P_2 = L^2 \langle |\psi({\bf r})|^4 \rangle $ of an eigenstate with {the} energy closest to zero are presented in SM~\cite{SuppMat}. In the localized phase, $P_2$ quickly saturates, when $L$
{exceeds}
$\xi_{\text{2D}}$. On the other hand, in the metallic phase, $P_2$ decreases with increasing $L$. In Fig.~\ref{fig:phase}, we show by a color code the IPR exponent $\tau_2(L) = - \partial \ln P_2(L) / \partial \ln L$ calculated in the range of our largest $L$. A nice agreement with the phase boundary obtained from the finite-size scaling analysis is observed.

\begin{figure}
	\centering
	\includegraphics[width =.39\textwidth]{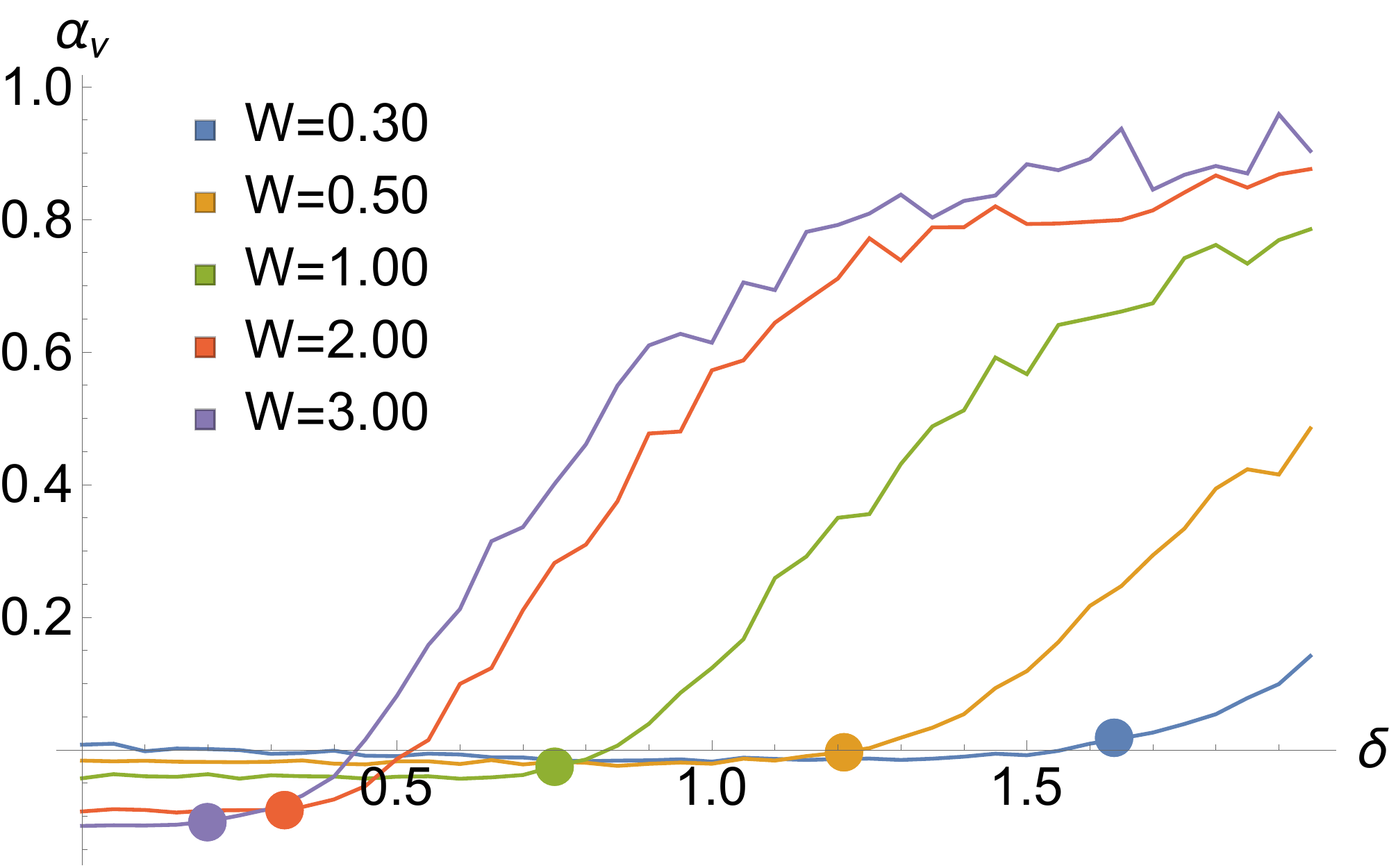}
	\caption{ Exponent $\alpha_\nu (W, \delta)$ of the DOS scaling, $\nu(\epsilon) \propto \epsilon^{\alpha_\nu}$, across the MIT at $W=0.3, 0.5, 1, 2, 3$. Positions of the MIT critical points $\delta_c(W)$ and the corresponding values $\alpha_\nu(W, \delta_c(W))$ are marked by dots.  }
	\label{fig:dos}
\end{figure}

{\it Density of states.}  In Fig.~\ref{fig:dos}, we show the exponent $\alpha_\nu (W, \delta)$ characterizing the scaling of the density of states (DOS)  $\nu(\epsilon) \propto \epsilon ^{\alpha_\nu}$ across the transition for various $W$.  In the metallic phase, $|\delta| < \delta_c(W)$, the RG predicts $\alpha_\nu \to -1$ at $L \to \infty$. The RG flow to this (infinite-randomness) fixed point is, however, logarithmically slow, which explains the observed non-universal values strongly different from $-1$,  see Table~\ref{tab:MF}. We also show there the related exponent $x_\nu = 2\alpha_\nu / (1+ \alpha_\nu)$ controlling the $L$ scaling of the DOS, $\nu(L) \sim L^{-x_\nu}$.  An apparent non-universality is observed also at criticality, $\delta = \delta_c$, see Table~\ref{tab:critical-parameters}; it is analogous to the corresponding property of { critical $\xi_M/M$ (discussed above) and $\sigma$.} When the system is driven into the localized phase by increasing $\delta$, we observe a power-law behavior with an exponent $\alpha_\nu$ growing and becoming positive, in consistency with previous fundings~\cite{motrunich2002particle-hole, bocquet2003network}.

\begin{table}
	\centering
	\begin{tabular}{c|cccccc}
		$W$ & $\alpha_\nu$ & $x_\nu$ &   $\sigma$ & $b$ & $\sigma_{\rm 1}$ & $\kappa_{\rm 1}$ \\
		\hline
		0.3  & \ $\sim -0.001$\  &  \ $\sim -0.002$\   &  \ 48.6 \   &\ 0.022 \  & \ 41.5\  & $\ \sim 4 \ $ \\
		0.5 & $-0.005$ & $-0.01$ &  25.8 & 0.036 & 27.8 & 7.7  \\
		1.0 & $-0.017$ & $-0.035$ &  9.8  & 0.10  & 9.1 & 2.8 \\
		2.0 & $-0.10$ & $-0.22$ &  4.4 & 0.24 & 4.2 & 3.2   \\
		3.0 & $-0.12$ & $-0.27$ &  3.6 & 0.30 & 3.3 & 2.4
	\end{tabular}
	\caption{Properties of the metallic phase ($\delta=0$, various $W$). Exponents $\alpha_\nu$ and $x_\nu$ of the DOS scaling, the conductivity $\sigma$ from transport calculation, and couplings $b$, $\sigma_1$, and $\kappa_1$ from a one-loop parabolic fit to the multifractality spectrum.
	}
	\label{tab:MF}
\end{table}

{\it Conductivity.}  We have further studied the conductivity $\sigma(L)$ at the transition and deep in the metallic phase. For this purpose, we evaluated the conductance $g(L,M)$
(measured in units of $e^2/h$) of a wide sample (width $M$ considerably exceeding the length $L$) using the Kwant software package~\cite{kwant}, see SM~\cite{SuppMat} for detail. The conductivity is then obtained as $\sigma(L) = \pi g(L,M) L/M$.  In the metallic phase, $\sigma(L)$ for sufficiently large $L$ is independent on $L$; the corresponding values for $\delta=0$ are given in Table~\ref{tab:MF}. The $L$-independence of $\sigma(L)$ holds also at criticality, $\delta = \delta_c$; these values are presented in Table~\ref{tab:critical-parameters}. 

{\it  Multifractality.} Moments of critical eigenfunctions exhibit multifractality, $L^2 \langle |\psi({\bf r})|^{2q} \rangle \sim L^{-\Delta_q}$. Equivalently, one can study multifractality of {the} local DOS, $\langle \nu^{q} ({\bf r}) \rangle \sim L^{-x_q}$; the two sets of exponents are related via $x_q = \Delta_q + q x_\nu$. For a chiral class (bipartite lattice), one can also define moments involving wavefunctions on nearby sites ${\bf r}$ and  ${\bf r'}$ belonging to different sublattices:
$L^2 \langle |\psi({\bf r})|^{2q}  |\psi({\bf r'})|^{2q'} \rangle \sim L^{-\Delta_{q,q'}}$  and
$\langle \nu^{q} ({\bf r}) \nu^{q'} ({\bf r'}) \rangle \sim L^{-x_{q,q'}}$, with $x_{q,q'} = \Delta_{q,q'} + (q+ q') x_\nu$. In the metallic phase, the multifractal exponents can be obtained in one-loop approximation controllable for large $\sigma$. In particular, one-loop results for $x_q$ and for sublattice-symmetric exponents $x_{q/2,q/2}$ read
\begin{eqnarray}
&& x_q \simeq b q(1-q) +x_\nu q^2\,; \ \ \ \Delta_q \simeq (b-x_{\nu}) q(1-q) \,,
\label{eq:Delta-q}
\\
&& x_{q/2,q/2} \simeq b q (1-q/2) \,, \label{eq:x-q2-q2}
\end{eqnarray}
with $b = 1/\sigma$ and $x_\nu= \kappa/\sigma^2$.
Our numerical results for the exponents $\Delta_q$ and $x_{q/2,q/2}$ in the metallic phase and at the MIT are presented in SM~\cite{SuppMat}. In the metallic phase, the data are well described by the one-loop form~\eqref{eq:Delta-q}  and~\eqref{eq:x-q2-q2}. The corresponding one-loop fit parameters $b$, $\sigma_1=1/b$, and $\kappa_1$ are shown in Table~\ref{tab:MF}. We emphasize an excellent agreement between $\sigma_1$ and Landauer conductivity $\sigma$.

At the MIT, the numerically obtained multifractality spectra deviate strongly from the parabolic form, which indicates violation (at least, partial) of the conformal invariance, as was also found for other 2D Anderson-transition points~\cite{karcher2021generalized, karcher2022generalized, karcher2022generalized-2}.
Furthermore, parameters of the multifractal spectra turn out to vary substantially along the critical line, which is another manifestation of the apparent non-universality discussed above.

\begin{figure}
	\centering
	\includegraphics[width=0.23\textwidth]{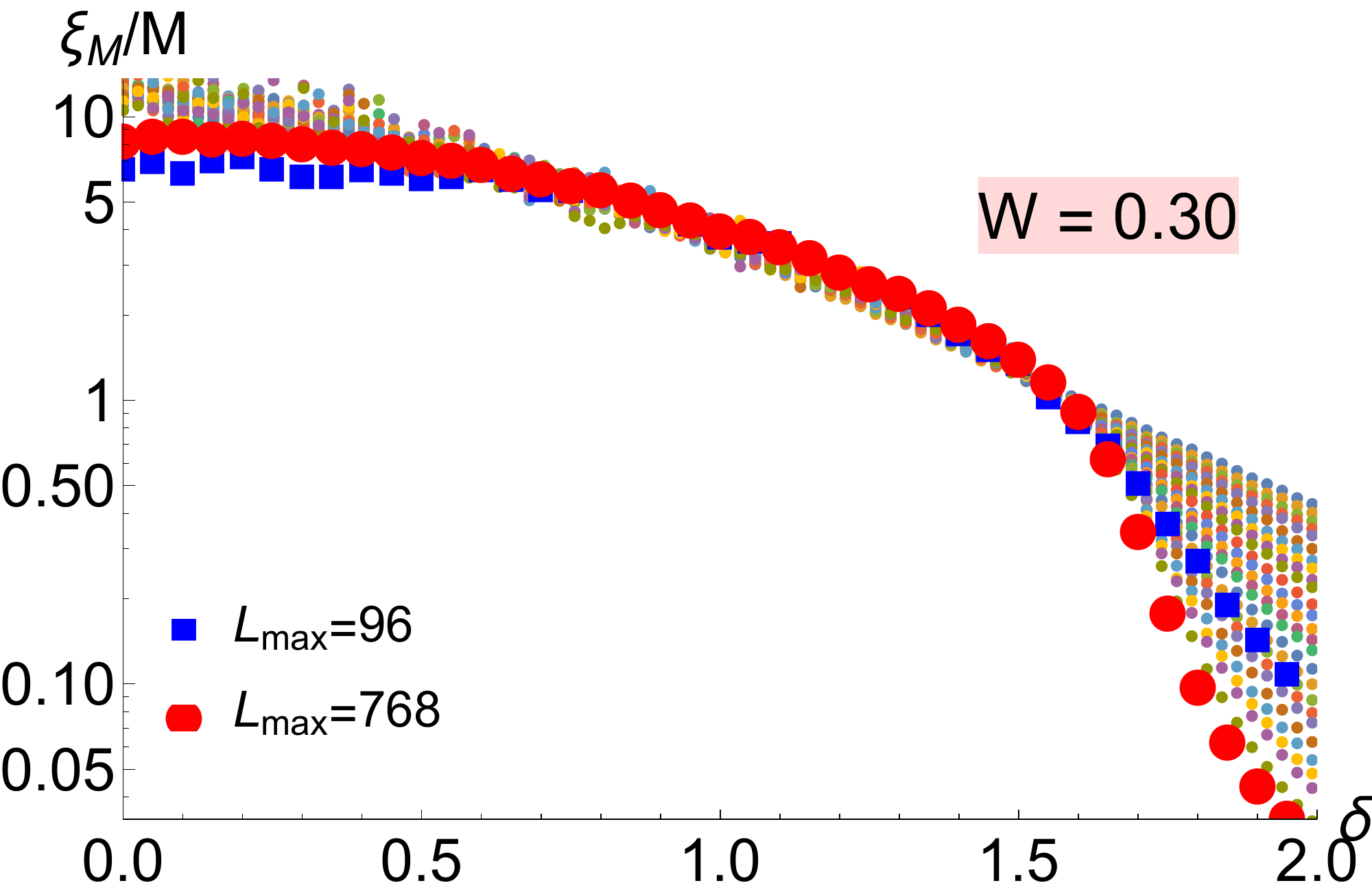}
	\includegraphics[width=0.23\textwidth]{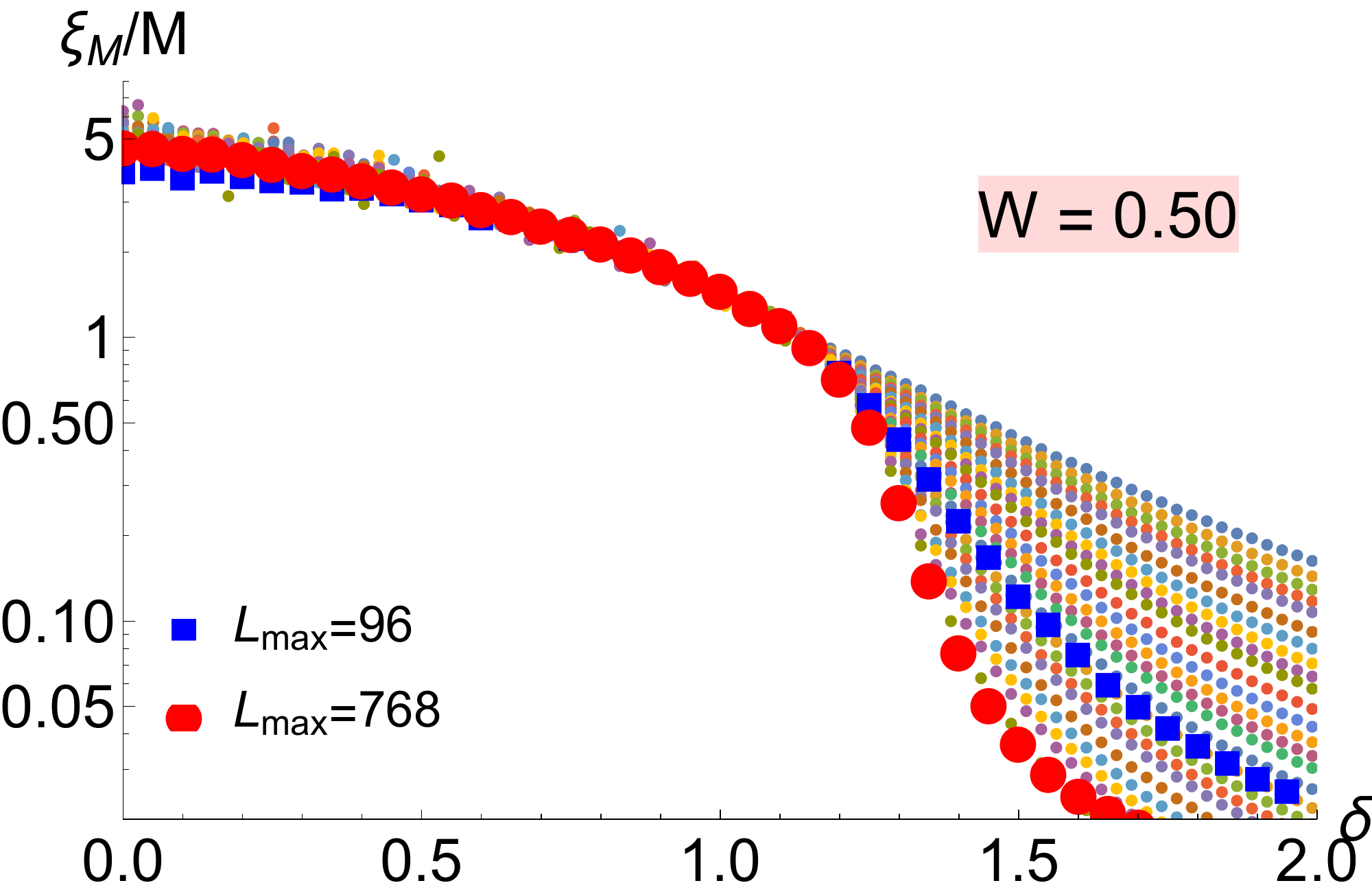}
	\includegraphics[width=0.23\textwidth]{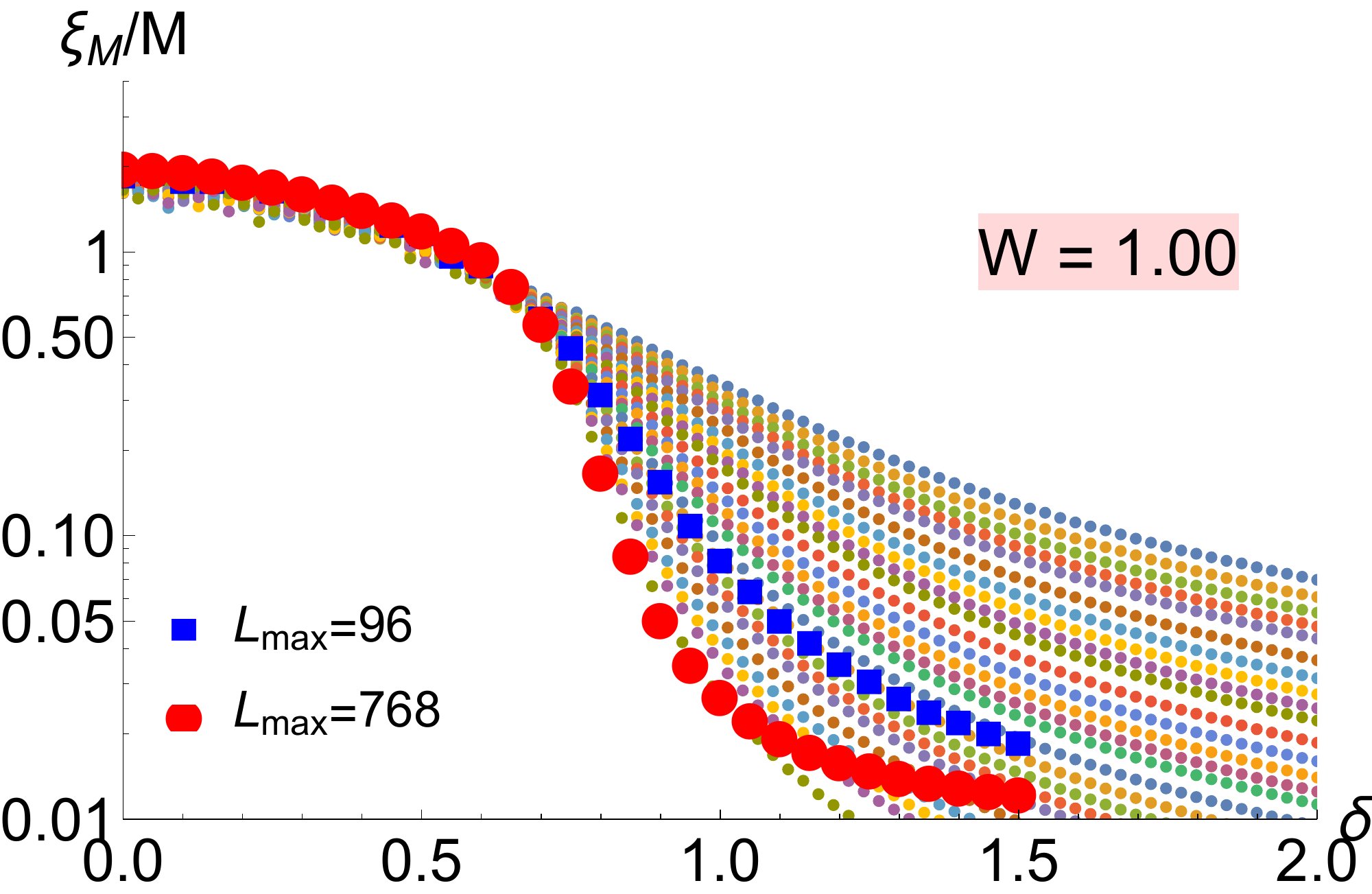}
	\includegraphics[width=0.23\textwidth]{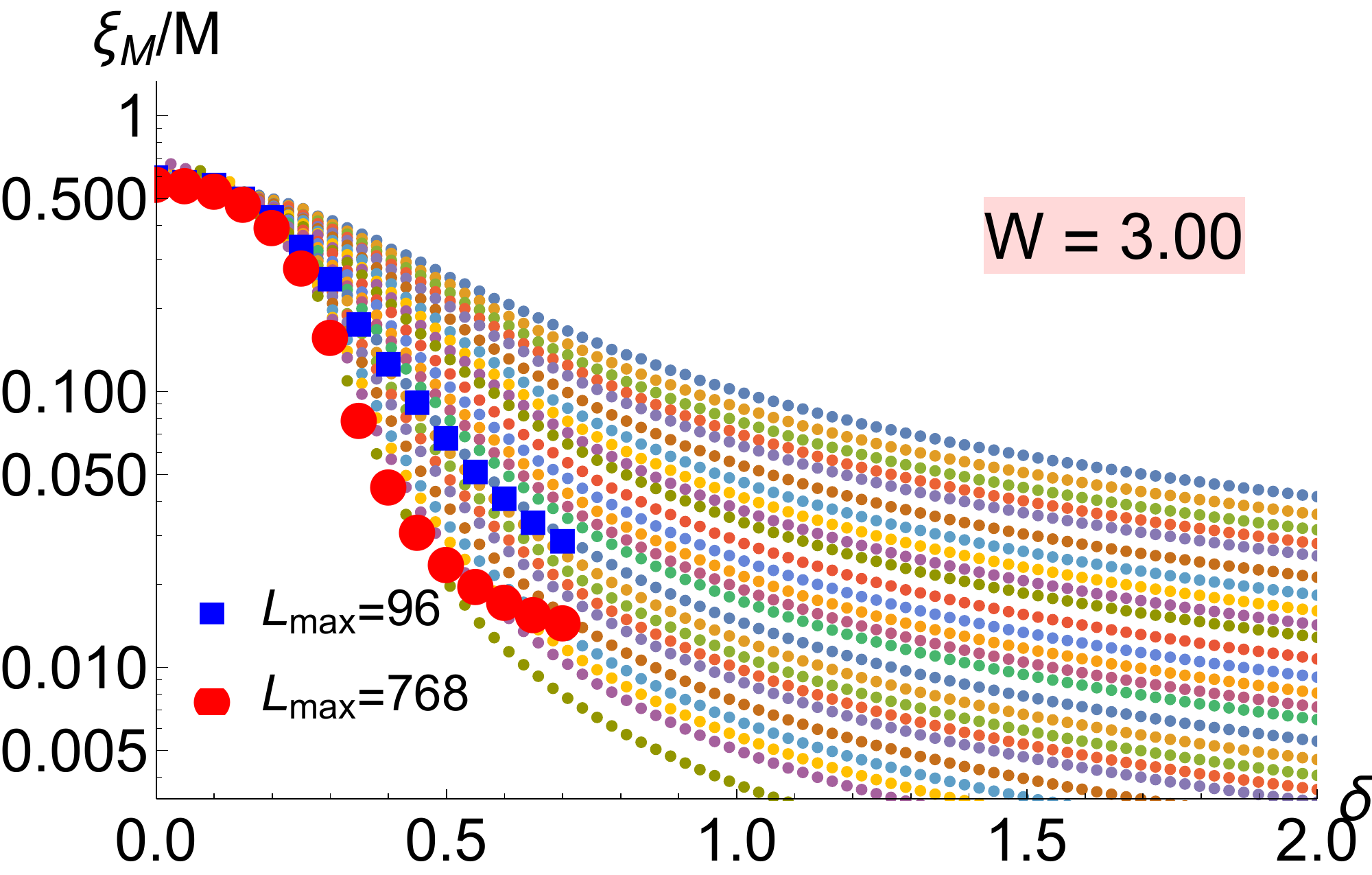}	
	\caption{ Small symbols: ratio $\xi_M/M$ from transfer-matrix analysis of a quasi-1D system with $W=0.3$, 0.5, 1, and 3  (cf. Fig.~\ref{fig:finite-size-scaling}). Large symbols:  $1/[2\pi (\alpha_0 -2 +x_\nu)]$ obtained by multifractal analysis of a 2D system  with $L \le 96$ (blue) and $L\le 768$ (red).
	The relation~\eqref{eq:exp-map} is fulfilled in the metallic phase and at criticality, see also Table~\ref{tab:critical-parameters}. 	
	 }
	\label{fig:transmf_aiii}
\end{figure}

{\it Quasi-1D to 2D conformal mapping.} An exponential map establishes a correspondence between a quasi-1D (infinite cylinder) and 2D (complex plane) geometries. Under the assumption of invariance of the critical theory under this conformal transformation, one can derive~\cite{Obuse-Conformal-2010} a relation (generalizing an earlier result of Ref.~\cite{Janssen-Statistics-1998})
\begin{align}
M/ \xi_M  &= 2\pi (\alpha_0 -2 +x_\nu) \,,
\label{eq:exp-map}
\end{align}
where $\alpha_0 = dx_q /dq |_{q=0}$. As shown in Fig.~\ref{fig:transmf_aiii} and in Table~\ref{tab:critical-parameters}, this relation indeed holds with a very good accuracy in our class-AIII model, both in the metallic phase and at {the} MIT.

{\it Summary and outlook.}   We have numerically studied the MIT in a 2D tight-binding system of class AIII by supplementing a quasi-1D finite-size scaling analysis with investigation of 2D conductivity, multifractality, and {the} DOS. The obtained phase diagram in the parameter plane of disorder $W$ and staggering $\delta$ is displayed in Fig.~\ref{fig:phase}.
Our findings agree with the sigma-model RG theory, with vortices driving a transition to the insulating phase~\cite{koenig2012metal-insulator}, yielding the flow shown in Fig.~\ref{fig:RG-flow}. We find $\nu = 1.55 \pm 0.1$ for the localization-length exponent, in agreement with the analytical estimate. Critical parameters at {the} MIT show an apparent non-universality, consistent with the analytically predicted slow renormalization along the critical line towards the ultimate $\sigma=0$ fixed point of the MIT. Non-parabolicity of the multifractal spectrum implies a violation of conformal invariance at the MIT. At the same time, our results support invariance with respect to the exponential conformal map between {the} cylinder and {the} plane geometries.

We foresee that future works will extend this investigation to (i) other models (e.g., on the hexagonal lattice) that are expected to provide access to strong-randomness fixed point of the MIT and (ii) to other chiral classes (BDI and CII). A detailed investigation of {the} generalized multifractality in the chiral classes will be published soon.

J.F.K. and A.D.M. acknowledge support by the Deutsche Forschungsgemeinschaft (DFG) via the grant MI 658/14-1.


\bibliography{gener-MF}

\pagebreak
\setcounter{equation}{0}
\setcounter{figure}{0}
\setcounter{table}{0}

\makeatletter
\renewcommand{\theequation}{S\arabic{equation}}
\renewcommand{\thefigure}{S\arabic{figure}}
\renewcommand{\thesection}{S-\Roman{section}}
\renewcommand{\bibnumfmt}[1]{[S#1]}
\begin{widetext}
	\setcounter{page}{1}
	\begin{center}
		Supplementary materials for \\
		\textbf{"Metal-insulator transition in a 2D system of chiral unitary class"}\\
		Jonas F.~Karcher$^{1,2}$, Ilya A.~Gruzberg$^{3}$, Alexander D.~Mirlin$^{1,2}$\\ 		
		$^{1}${Institute for Quantum Materials and Technologies, Karlsruhe Institute of Technology, 76021 Karlsruhe, Germany}\\
		$^{2}${Institut f\"ur Theorie der Kondensierten Materie, Karlsruhe Institute of Technology, 76128 Karlsruhe, Germany}\\
		$^{3}${Ohio State University, Department of Physics, 191 West Woodruff Ave, Columbus OH, 43210, USA}
	\end{center}
\end{widetext}
\onecolumngrid

These supplementary materials contain a detailed overview on finite size scaling analysis, numerical study of the scaling of the inverse participation ratio, exact diagonalization calculation of the density of states, determination of the conductivity and conductance using the transfer matrix. and multifractal analysis of LDOS moments.

\section{Finite-size scaling}
\label{sup:fss}

We follow the approach from Ref.~\cite{slevin2014critical} in order to find the critical staggering $\delta_c$ and the exponents $\nu$ and $y$ of our chiral model. We briefly sketch this formalism below. The dimensionless ratio $\xi_M/M$ is expressed in the scaling form
\begin{align}
\dfrac{\xi_M}{M} = F(\phi_1,\phi_2) \,,
\end{align}
where $F$ is a universal scaling function and $\phi_i$ are the scaling variables
\begin{align}
\phi_i&= u_i(w) M ^{\alpha_i}, &
u_i(w) &= \sum_{j=0}^{m_i} b_{i,j}w^j, & w&=\dfrac{\delta-\delta_c}{\delta_c}\,.
\end{align}
Here the reduced staggering $w$ is a control parameter that drives the systems through the MIT; we focus on the localized side, $w > 0$.
Further, $\phi_1$ is the relevant variable; the corresponding exponent $\alpha_1$ is related to the localization length exponent via $\alpha_1 = \nu^{-1}$. The second variable $\phi_2$ is the irrelevant one; the corresponding exponent is negative: $\alpha_2=-y_{\rm irr} < 0$. To simply the numerical optimization, one also expands the scaling function $F$ and truncates the expansion at a finite order:
\begin{align}
F&= \sum_{j_1 = 0}^{n_1} \sum_{j_2 = 0}^{n_2} a_{j_1,j_2}\phi_1^{j_1}\phi_2^{j_2},
\end{align}
where $a_{1,0}$ and $a_{0,1}$ are set to unity in order to avoid ambiguity of the fitting model.
The total number of parameters in the fitting model is
\begin{align}
N_{\rm p} &= 2+ m_1+m_2 +(n_1+1)(n_2+1) \,.
\end{align}
In the present work , we fix $m_1=1$, $m_2=0$, $n_1 = 2$, and $n_2=1$, which allows us to
avoid fitting with too many parameters.

\begin{figure}
	\centering
	\includegraphics[width=.45\linewidth]{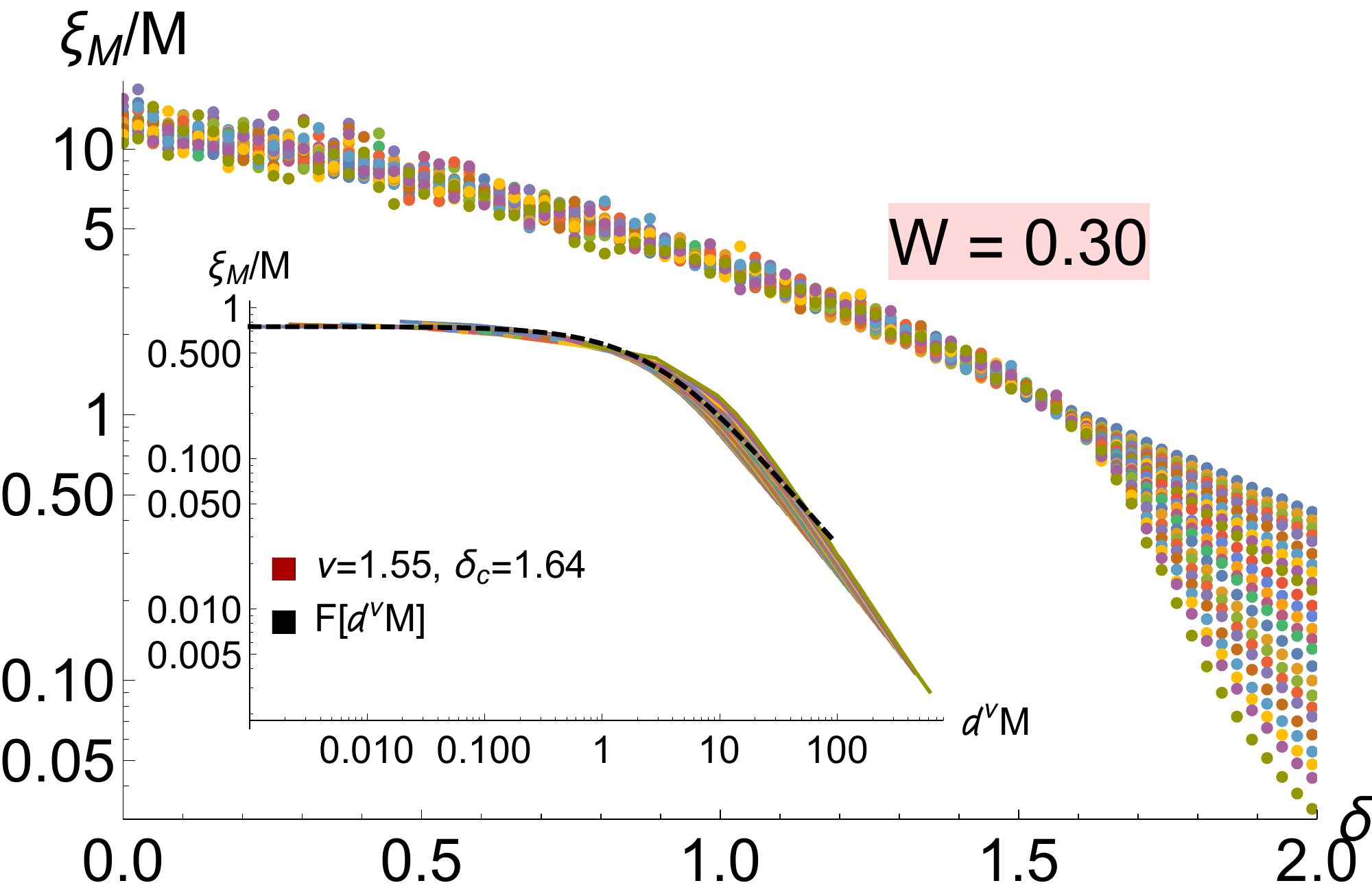} \ \ \
	\includegraphics[width=.45\linewidth]{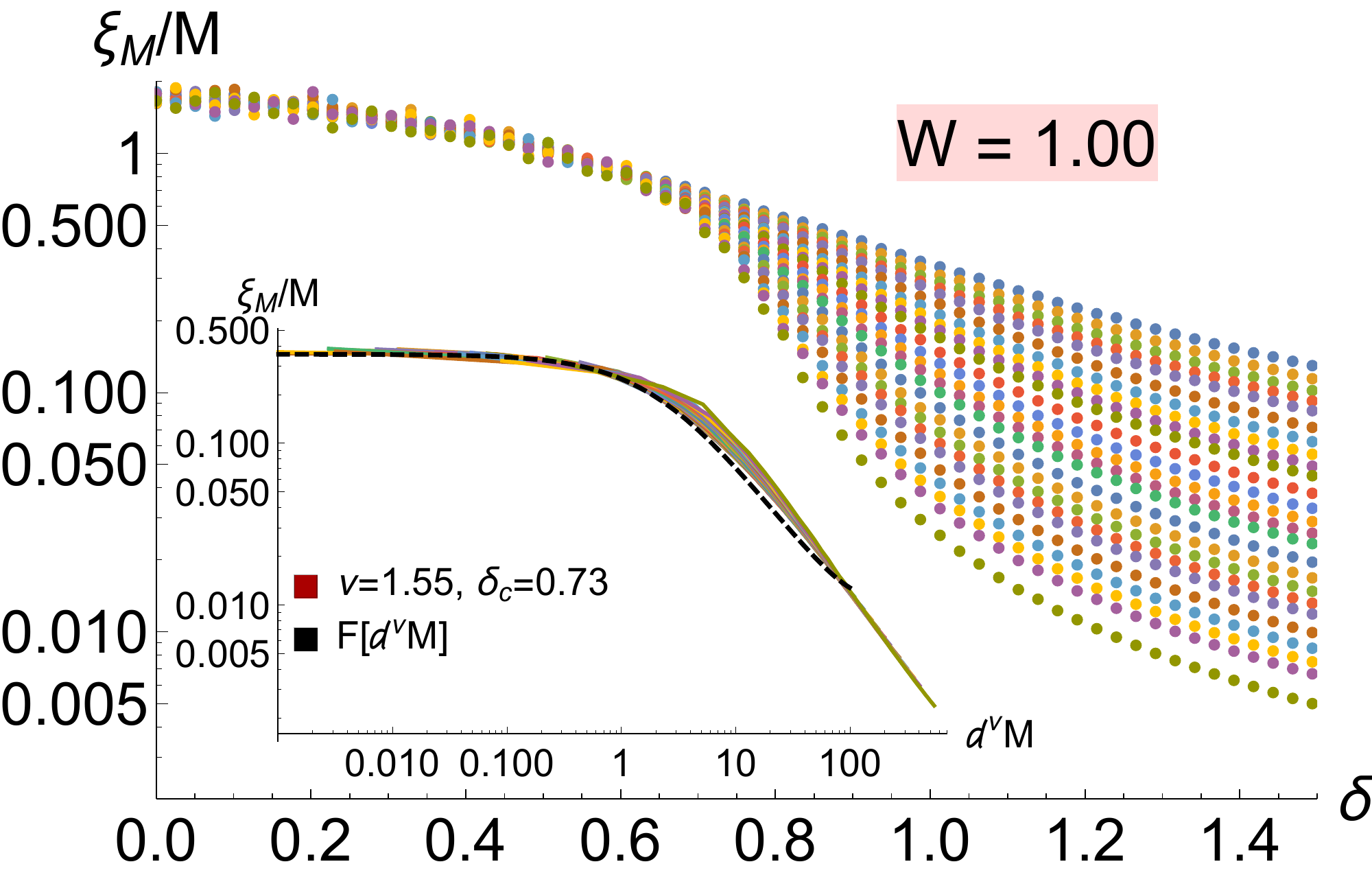} \\[0.3cm]
	\includegraphics[width=.45\linewidth]{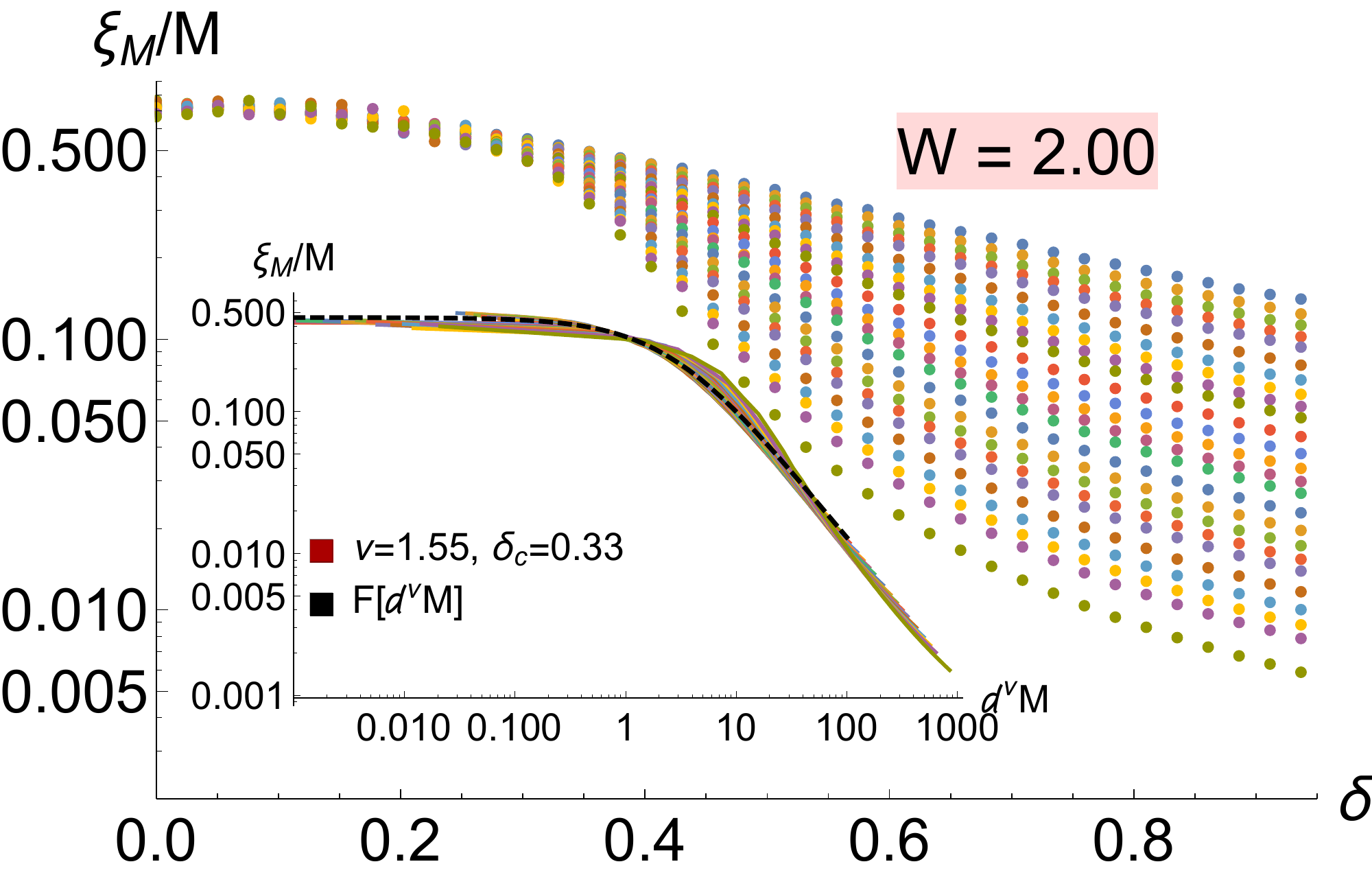} \ \ \
	\includegraphics[width=.45\linewidth]{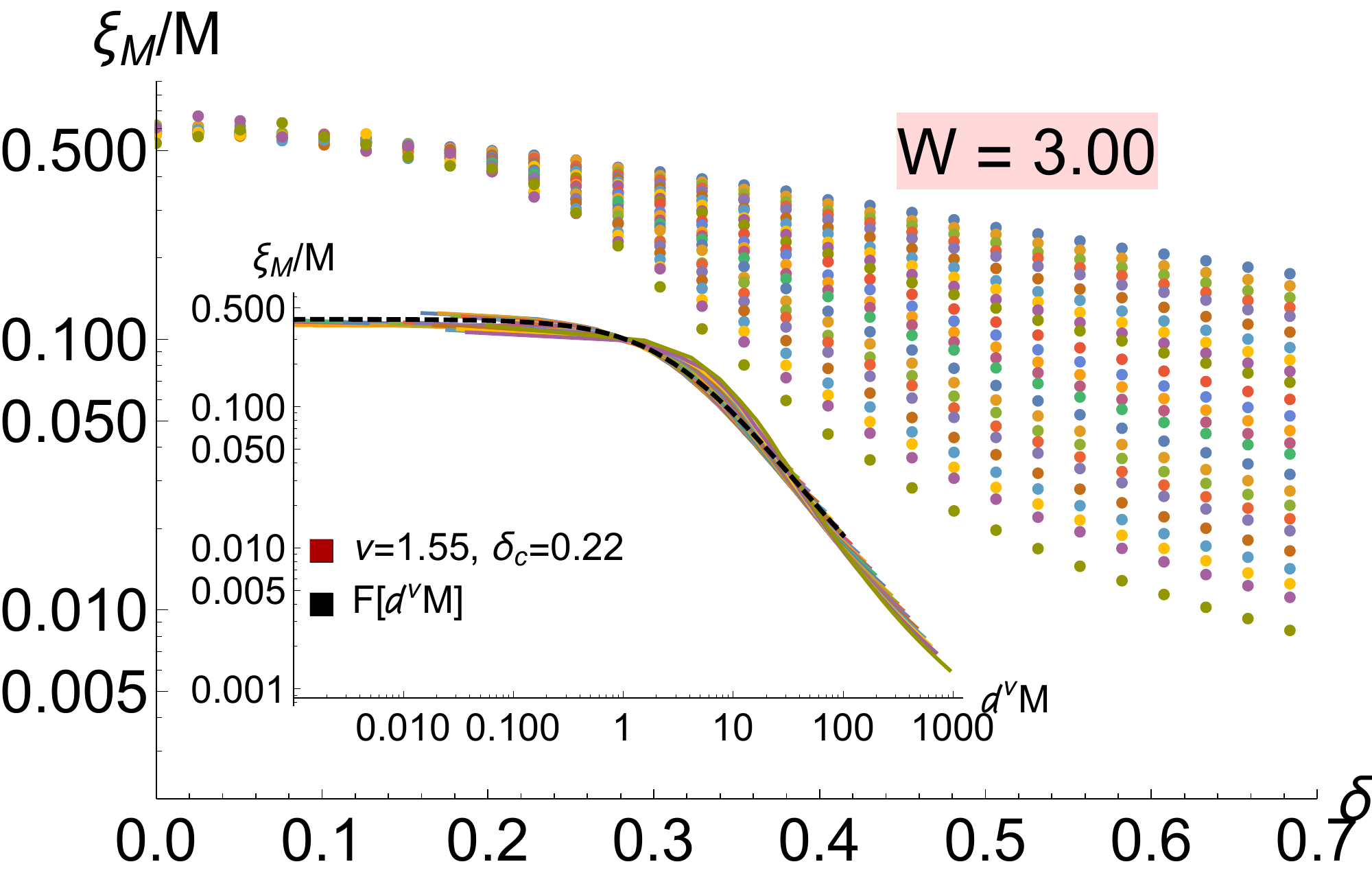}
	\caption{ Finite-size scaling analysis. The ratio $\xi_M/M$ as a function of the staggering $\delta$ for disorder strengths $W=0.3, 1.0, 2.0, 3.0$ and $M=12, \ldots,256$. Insets: data collapse $\xi_M/M = F(d^\nu M)$, with $d = \delta-\delta_c$, and the exponent $\nu =1.55$.  The critical staggering values $\delta_c$ are indicated in the legends; see also Table~\ref{tab:critical-parameters} of the main text.}
	\label{fig:fit_aiiis}
\end{figure}

We consider the values of disorder $W=0.3, 0,5, 1.0, 2.0, 3.0$. For each of them, we perform the analysis in a wide interval of the staggering $\delta$ by
calculating $\xi_M/M$ for many system sizes $M$ in the range $M = 12,\ldots, 256$ using the length $L=10^5$.
 The best fits and collapses of the data are shown in Fig.~\ref{fig:finite-size-scaling} of the main text for $W=0.5$ and
 in Fig.~\ref{fig:fit_aiiis} for the disorder strengths $W=0.3, 1.0, 2.0, 3.0$.

\section{Inverse participation ratio}
\label{sup:ipr}
In the localized phase, the inverse participation ratio (IPR) $P_2(L)$ quickly saturates when $L$ exceeds the localization length $\xi_{2\rm D}$. At criticality or in the metallic phase, $P_2(L)$ decreases without bound as $L$ increases. Therefore, the IPR and the effective IPR exponent $\tau_2(L)$,
\begin{align}
P_2(L) &= L^2\left\langle |\psi({\rm} r)|^4\right\rangle\,, & \tau_2(L) &= -\dfrac{\partial \ln P_2(L)}{\partial\ln L} \,,
\label{eq:ipr}
\end{align}
are good indicators for metallic or insulating behavior at a given length scale.

\begin{figure}
	\centering
	\includegraphics[width=0.45\textwidth]{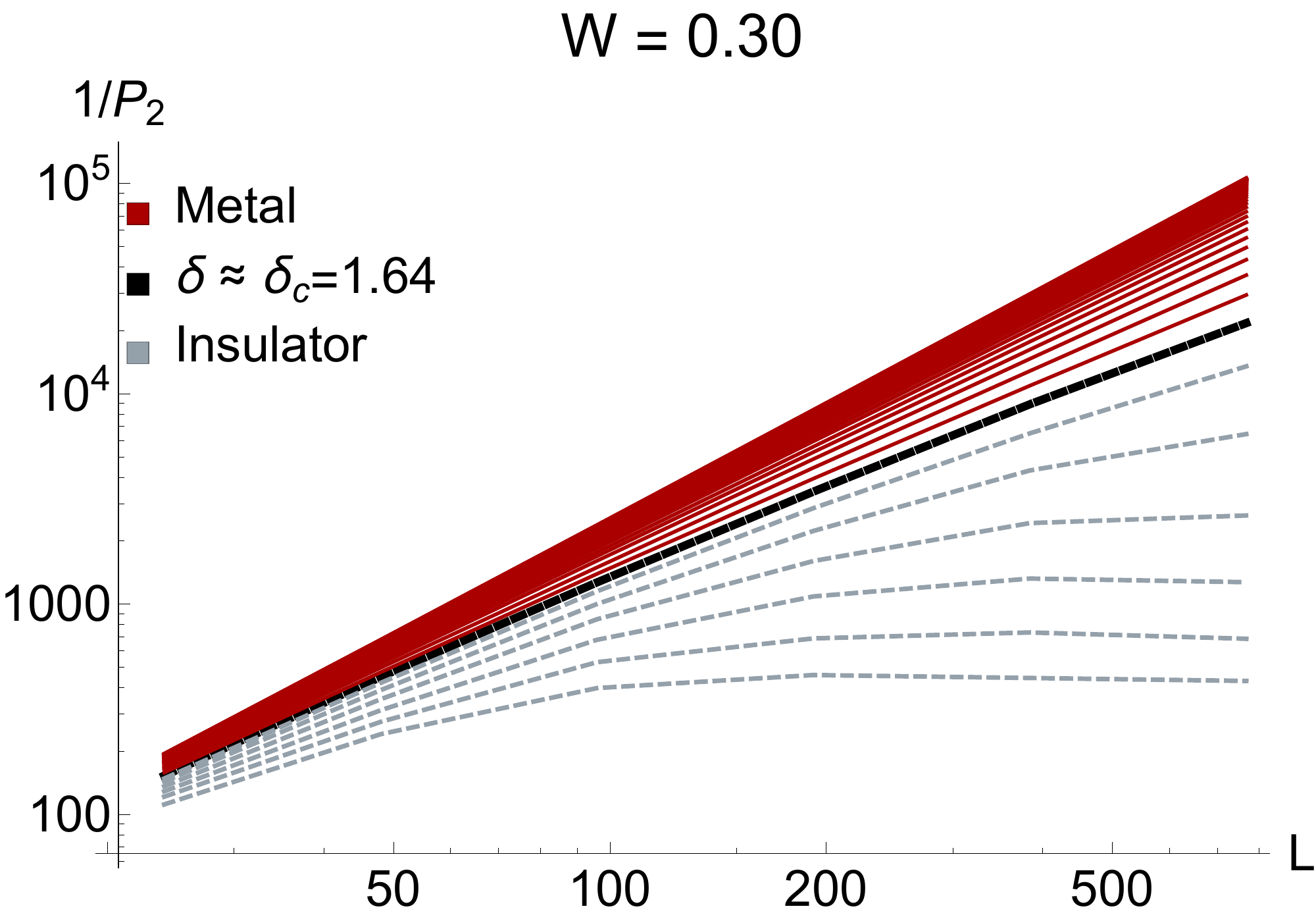} \ \ \
	\includegraphics[width=0.45\textwidth]{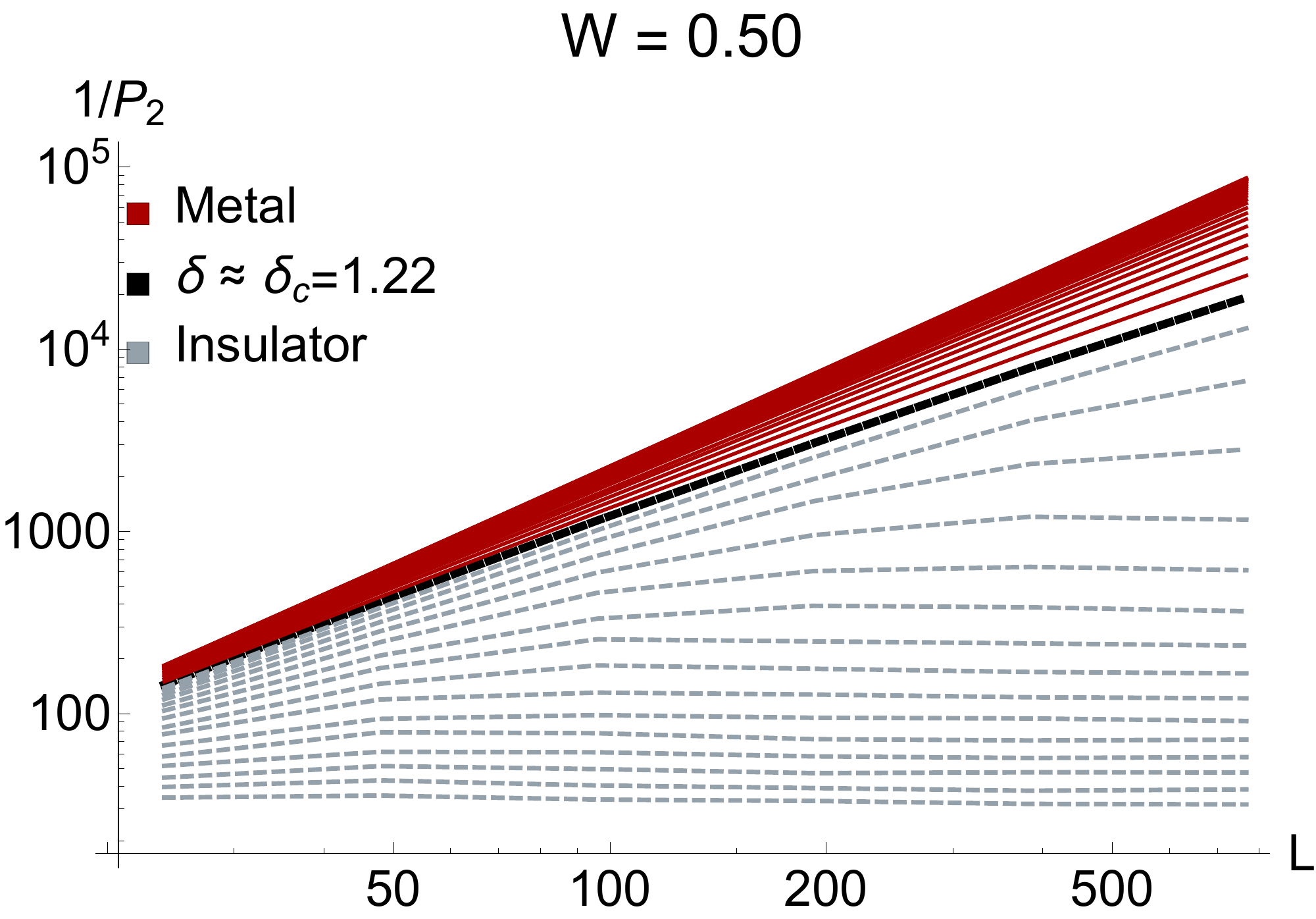} \\[0.3cm]
	\includegraphics[width=0.45\textwidth]{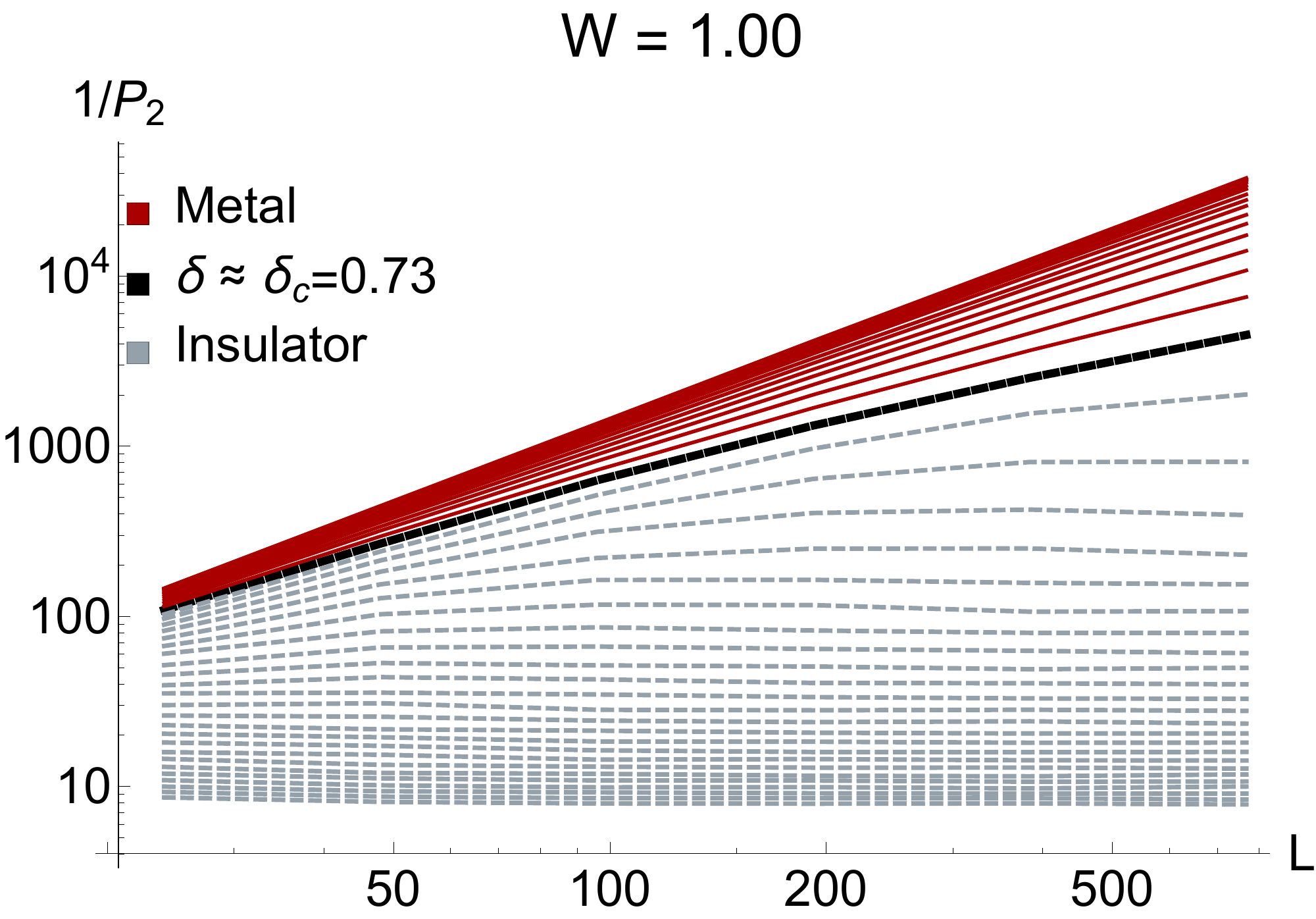} \ \ \
	\includegraphics[width=0.45\textwidth]{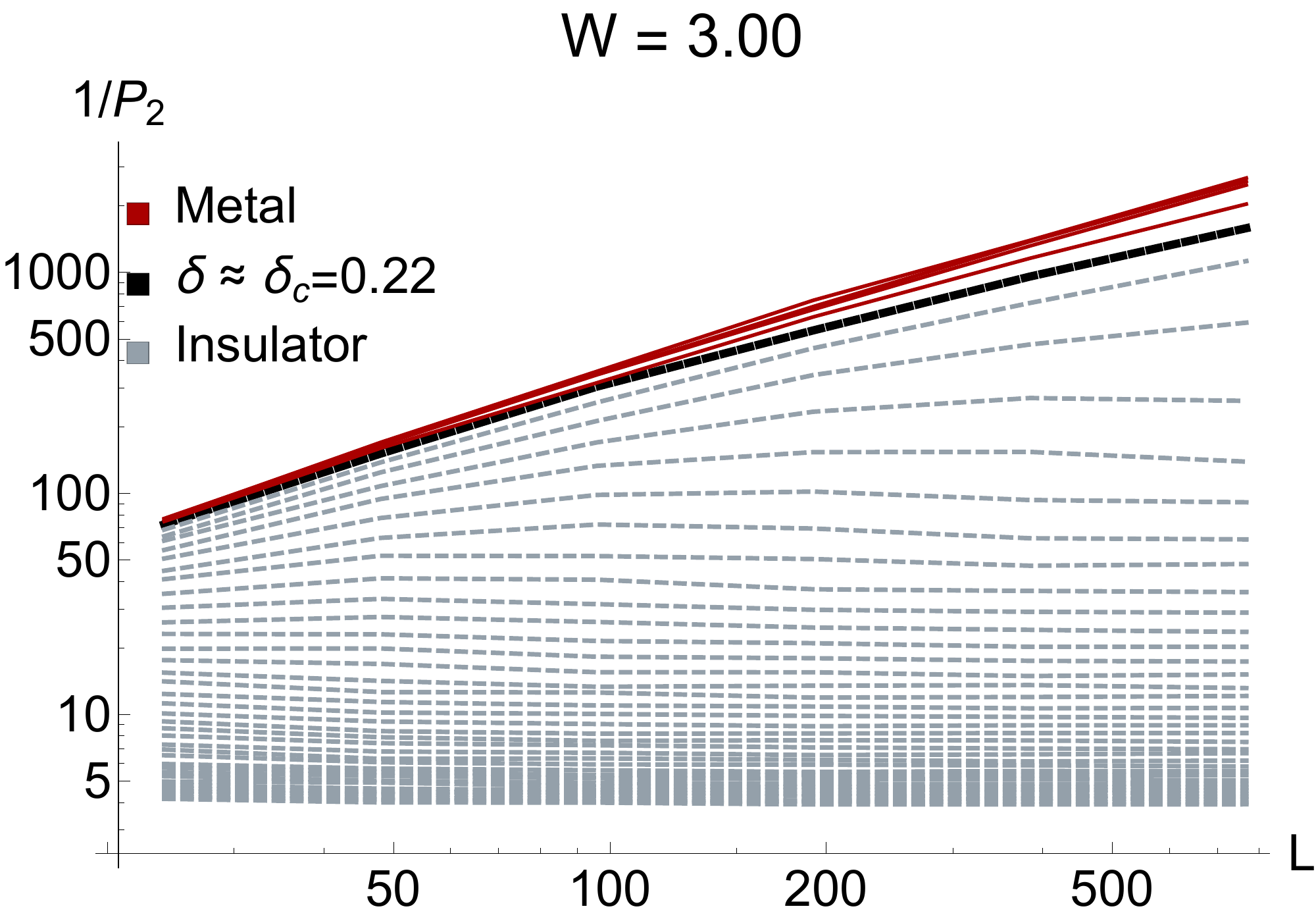}	
	\caption{ Scaling of the IPR $P_2$ across the transition.  Curves show $1/P_2(L)$ for square $L \times L$ samples for disorder strengths $W=0.3, 0.5, 1.0$, and 1.3, and for a range of staggering $\delta$ around the critical value $\delta_c(W)$. The upper curve in each panel corresponds to $\delta=0$, which is in the metallic phase ($1/P_2(L)$ keeps increasing), while the lower curves correspond to values of $\delta$ considerably exceeding $\delta_c$, which are thus well in the insulating phase ($1/P_2(L)$ saturates). The curve corresponding to $\delta$ that is close to the critical value $\delta_c$ is highlighted. }
	\label{fig:pr_aiii}
\end{figure}

In Fig.~\ref{fig:pr_aiii}, we show the scaling of $1/P_2(L)$ for disorder strengths $W=0.3, 0.5, 1.0, 3.0$ and for a range of staggering $\delta$ around the critical value $\delta_c(W)$ (starting with $\delta=0$, which is in the metallic phase, up to values of $\delta$ considerably exceeding $\delta_c$, which are thus well in the insulating phase).
A transition form the metallic to the insulating behavior is manifest. One can estimate the position of the MIT by studying the minimal staggering at which the participation ratio $P_2(L)^{-1}$ stops diverging. For this purpose, it is instructive to consider the effective exponent $\tau_2(L)$, which rapidly converges to zero in the insulating phase for $L>\xi_{2\rm D}$. Numerical data for $\tau_2(L)$ in large systems ($L=768$) is shown by color code in the phase diagram in Fig.~\ref{fig:phase} of the main text. A full consistency with the phase boundary determined by the finite-size scaling approach is observed.

\section{Density of states}
\label{sup:dos}

The density of states scales generically as a power law at low energies.
The corresponding exponent $\alpha_\nu$  is directly related to the anomalous dimension $x_\nu$ of the operator that is coupled to the energy:
\begin{align}
\langle \nu(\epsilon) \rangle &\propto \epsilon^{\alpha_\nu}, 
& 
\langle \nu(L) \rangle &\propto L^{-x_\nu} 
& 
x_\nu&= \dfrac{2\alpha_\nu}{1+\alpha_\nu}.
\label{eq:dos_l}
\end{align}
We determine $\alpha_\nu$ numerically as a function of the staggering $\delta$ and the disorder strength $W$ in a large region of the phase diagram.

\begin{figure}
	\centering
	\includegraphics[width=0.31\textwidth]{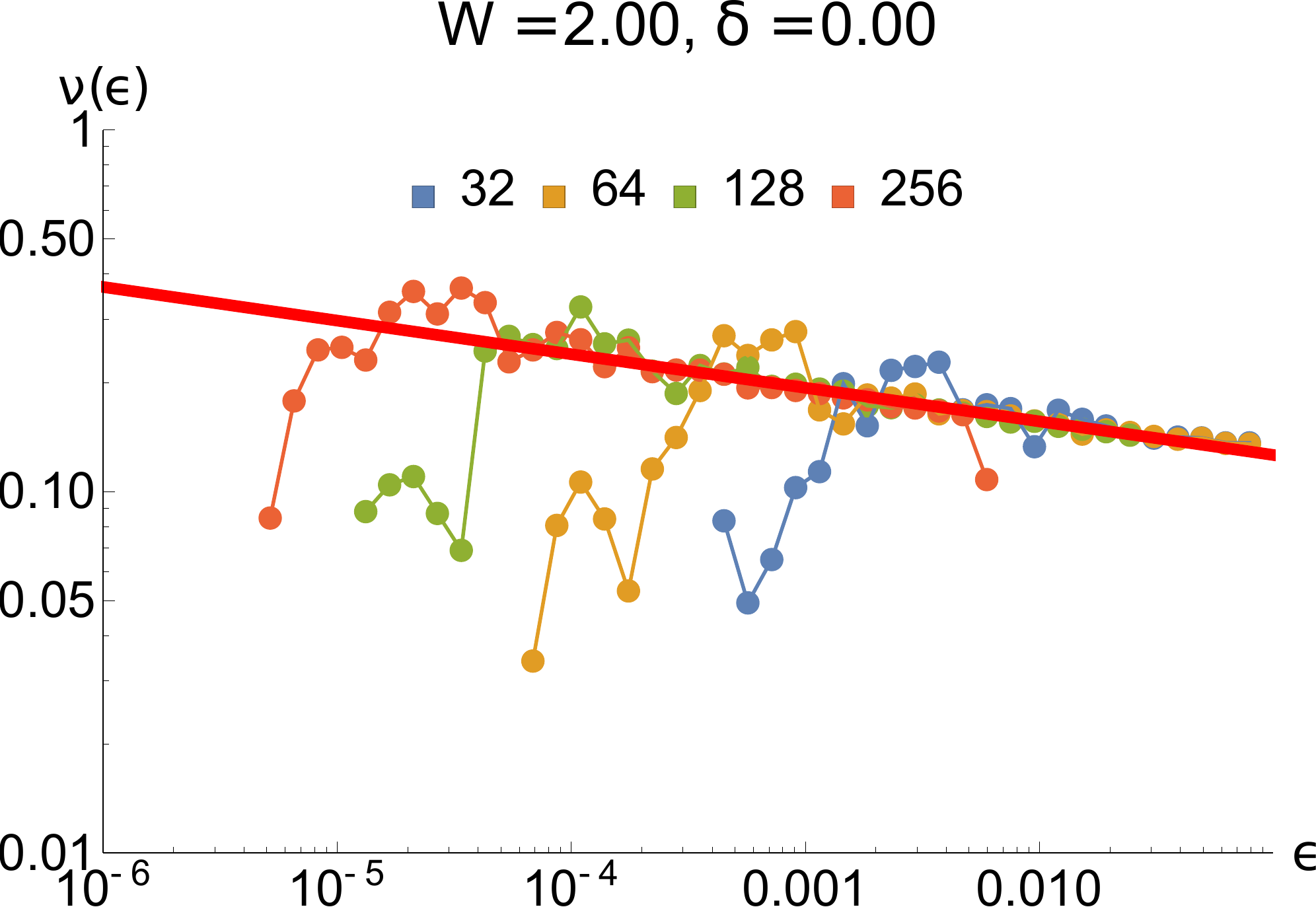} \ \ \
	\includegraphics[width=0.31\textwidth]{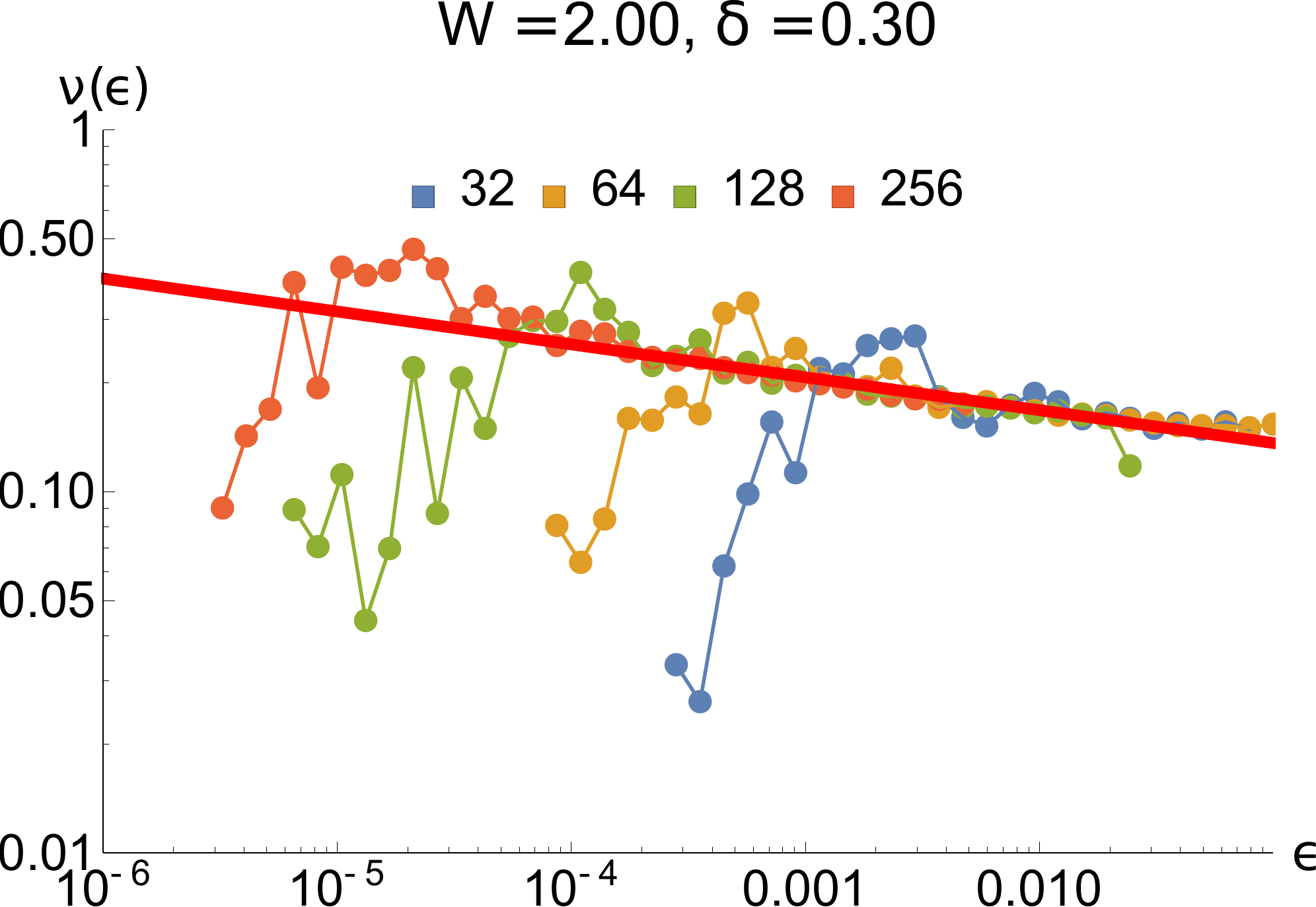} \ \ \
	\includegraphics[width=0.31\textwidth]{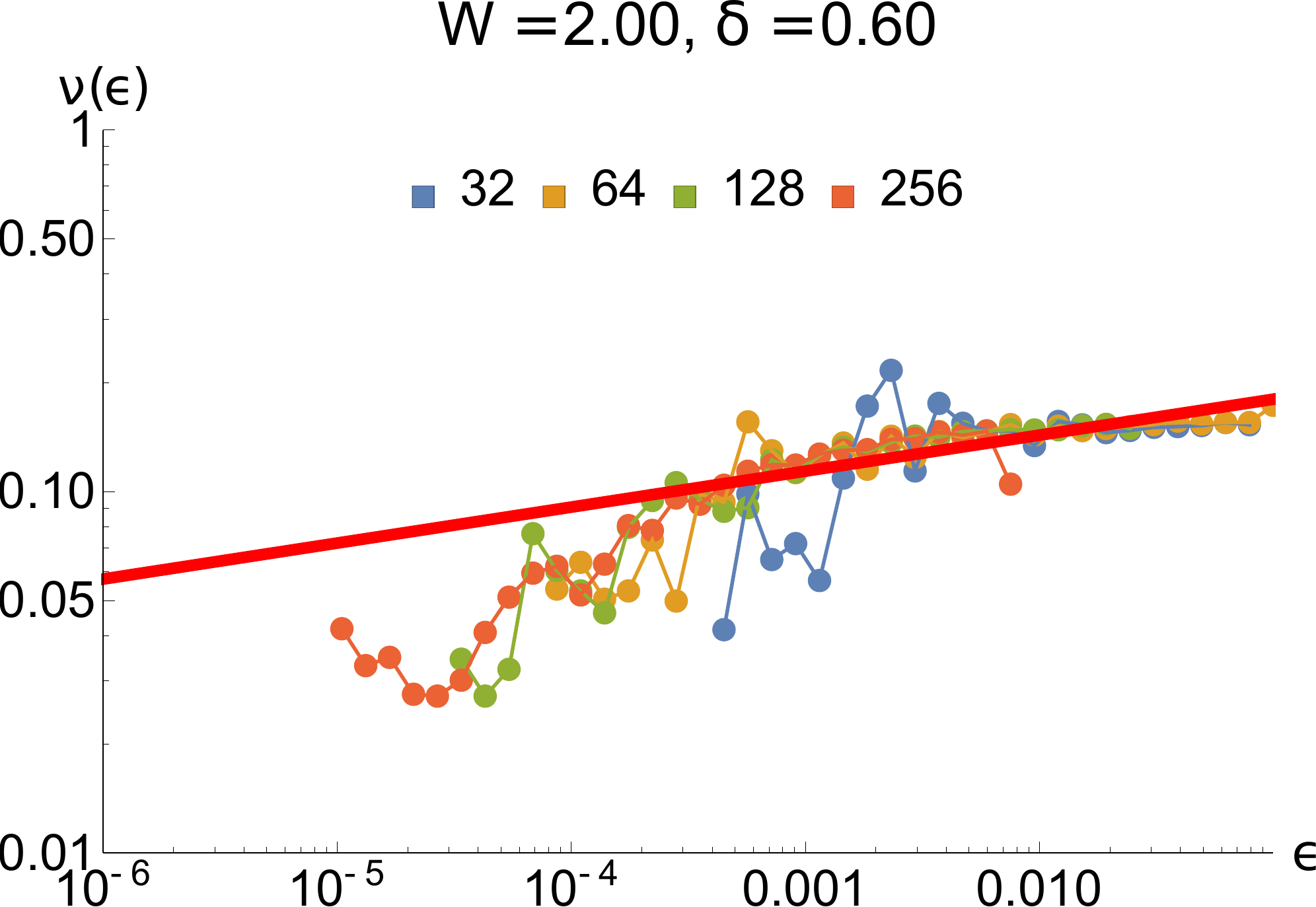}
	\caption{The density of states $\nu(\epsilon)$ of the chiral Hamiltonian~\eqref{eq:ham_aiii} for the disorder strength $W=2.0$. Each panel shows data for four system sizes, $L= 32, 64, 128$, and 256.	{\it Left panel:} No staggering $\delta=0$; the system is deeply in the metallic phase. The  exponent $\alpha_\nu<0$, i.e., the DOS diverges at $\epsilon \to 0$. On each of the curves, RMT oscillations are observed for the lowest energies (of the order of the energy of the lowest state in the system of the corresponding size). {\it Middle panel:} $\delta=0.3$, which is close to the critical value $\delta_c$. {\it Right panel:}  Strongly insulating regime, $\delta=0.6$. Here the exponent $\alpha_\nu>0$, i.e., the DOS vanishes in the limit $\epsilon\to 0$. }
	\label{fig:doss}
\end{figure}

For this purpose, the exact diagonalization of the chiral Hamiltonian~\eqref{eq:ham_aiii} is carried out for each set of parameters, for system sizes $L=32, 64, 128, 256$, and for $10^4$ disorder configuration. For each configuration, the 64 smallest eigenvalues are binned into $50$ intervals in the energy range $[10^{-6},1]$ in order to find an approximation for the DOS. This is illustrated in Fig.~\ref{fig:doss}, where the data for $W=2.0$ and for three values of the staggering $\delta$ are shown: (i) deep in the metallic phase, (ii) close to criticality, and (iii) in the strong insulator limit. We fit the exponent $\alpha_\nu$ in the energy interval where a power law behavior holds with a good approximation. The resulting exponents are shown in Fig.~\ref{fig:dos} and in Tables ~\ref{tab:critical-parameters} and~\ref{tab:MF} of the main text.   They are also used in the evaluation of the right-hand side of Eq.~\eqref{eq:exp-map} shown in Fig.~\ref{fig:transmf_aiii} and in Table~\ref{tab:critical-parameters}.

\section{Conductivity}
\label{sup:cond}
We numerically calculated the conductivity $\sigma(L)$ deep in the metallic phase (at zero staggering) and at criticality by computing the conductance $g (L,M)$ of wide strips (aspect ratio $r = L/M \ll 1$) with the help of the Kwant software package~\cite{kwant}.
Specifically, the conductivity is given by $\sigma(L) = \lim_{M\to \infty} \pi g (L,M) L/M$.  (There is a factor $\pi$ here since we measure the dimensionless conductivity $\sigma$ in units of $e^2/\pi h$, while the dimensionless conductance is measured in units of $e^2/h$.) In practice, we use the aspect ratios $r=\frac{1}{10}, \ldots, \frac15, \frac14, \frac13$ to check that the resulting conductivity $\sigma_r(L)$ is essentially independent of $r$, thus approximating the true limiting value $\sigma(L)$.
We used $L=48,\ldots, 112$ and averaged the results over at least 30 disorder configurations. To our satisfaction, the conductivity $\sigma(L)$ in the metallic phase (at zero staggering, $\delta=0$) becomes independent of $L$ at large $L$, as expected. An analogous behavior is observed also at criticality, $\delta= \delta_c$. In fact, the RG theory predicts a very slow drift of the conductivity with the system size, $\sigma(L) \propto L^{-1/16}$, at the MIT critical point, see the main text. However, a reliable observation of this drift is very difficult in view of the limits on the range of available $L$ and would require much more computational time to further reduce statistical errors. At the same time, the apparent non-universality of the critical conductivity obtained in our simulations is in full consistency with these analytical predictions.

The obtained values of the conductivity $\sigma$ at MIT ($\delta = \delta_c(W)$) and in the metallic phase ($\delta=0$) are presented in Tables ~\ref{tab:critical-parameters} and~\ref{tab:MF} of the main text, respectively.

\section{Multifractality}
\label{sup:mult}

Moments of critical eigenfunctions exhibit multifractality. Equivalently, one can study multifractality of the local DOS:
\begin{align}
L^2 \langle |\psi({\bf r})|^{2q} \rangle &\sim L^{-\Delta_q}, & \langle \nu^{q} ({\bf r}) \rangle &\sim L^{-x_q}.
\end{align}
The two sets of multifractal exponents are related via $x_q = \Delta_q + q x_\nu$.
For a chiral class (bipartite lattice), one can also define moments involving wave functions on nearby sites ${\bf r}$ and  ${\bf r'}$ belonging to different sublattices:
\begin{align}
L^2 \langle |\psi({\bf r})|^{2q}  |\psi({\bf r'})|^{2q'} \rangle &\sim L^{-\Delta_{q,q'}}, & \langle \nu^{q} ({\bf r}) \nu^{q'} ({\bf r'}) \rangle &\sim L^{-x_{q,q'}},
\end{align}
with $x_{q,q'} = \Delta_{q,q'} + (q+ q') x_\nu$.

In the parabolic approximation, the multifractal exponents are determined only by two parameters $b$ and $x_\nu$:
\begin{eqnarray}
&& x_q \simeq b q(1-q) +x_\nu q^2\,; \qquad \Delta_q \simeq (b-x_{\nu}) q(1-q) \,,
\label{eq:Delta-qs}
\\
&& x_{q/2,q/2} \simeq b q (1-q/2) \,, \label{eq:x-q2-q2s}
\end{eqnarray}
Here $x_\nu$ is the exponent controlling the scaling of the average DOS, see Eq.~\eqref{eq:dos_l}. Equations~\eqref{eq:Delta-qs} and~\eqref{eq:x-q2-q2s} can be derived in the metallic phase by using the one-loop approximation (controlled for large conductivity $\sigma$) within the RG framework. In this approximation the parameters are $b = 1/\sigma$ and $x_\nu= \kappa/\sigma^2$.

The spectrum $x_{q/2,q/2}$ satisfies the symmetry $q\rightarrow 2-q$. It can be shown that this symmetry is exact (i.e. holds also beyond the parabolic approximation); it is a manifestation of a broader class of Weyl symmetries satisfied by the generalized multifractal exponents. A proof of this statement and a detailed investigation of the generalized multifractality in chiral classes will be published elsewhere. At the same time, the symmetry $q\rightarrow 1-q$ satisfied by $\Delta_q$ in Eq.~\eqref{eq:Delta-qs} is approximate, i.e., it does not generically hold (for the considered class AIII) beyond the parabolic approximation.

\begin{figure*}
	\centering
	\includegraphics[width=1.0\textwidth]{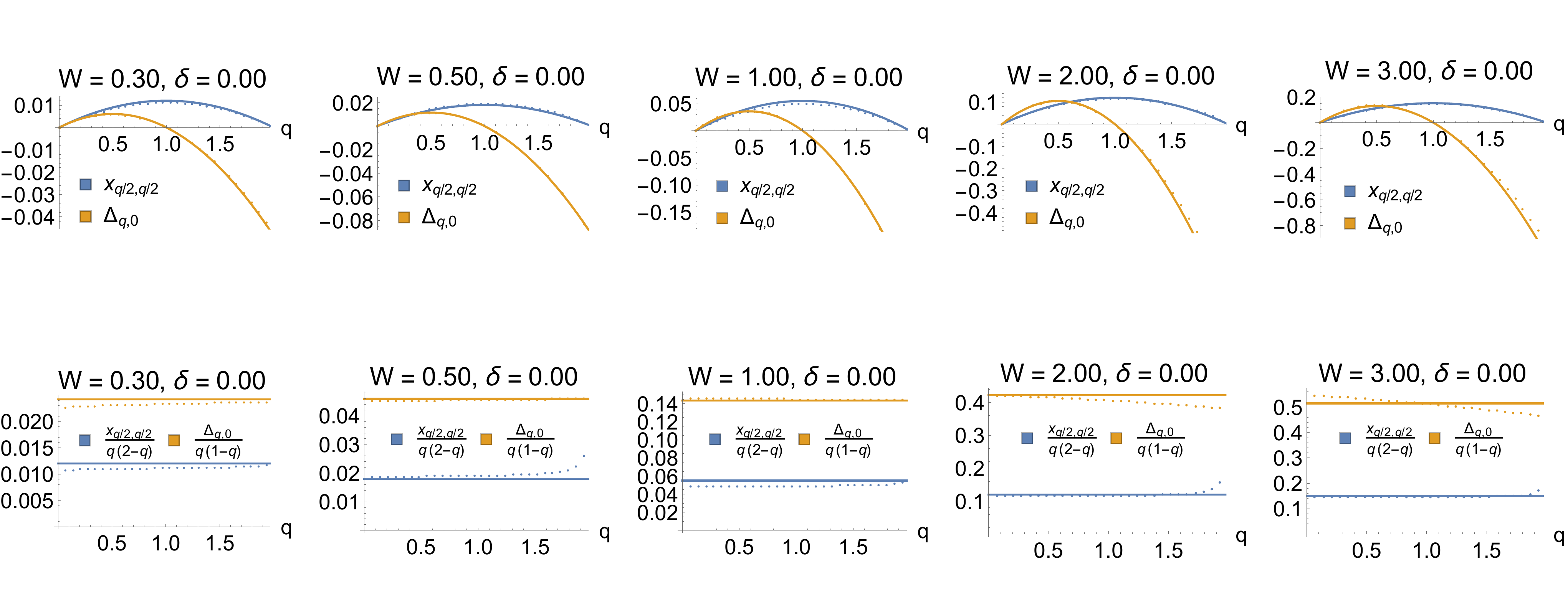}
	\caption{{\it Upper panels:} Multifractal exponents $\Delta_q$ (yellow dots) and $x_{q/2,q/2}$ (blue dots) as functions of $q$ in the metallic phase (no staggering, $\delta=0$) for disorder strengths $W=0.3, 0.5, 1.0, 2.0, 3.0$.  The solid curves are the parabolic approximations~\eqref{eq:Delta-qs},~\eqref{eq:x-q2-q2s}.
	{\it Lower panels:} Same data presented as $\Delta_q/[q(1-q)]$  and $x_{q/2,q/2}/[q(2-q)]$, respectively. In this representation, the parabolic approximations become horizontal straight lines.}
	\label{fig:mf-metal}
\end{figure*}

In Fig.~\ref{fig:mf-metal}, we show the multifractal spectra $\Delta_q$ and $x_{q/2,q/2}$ in the metallic phase (at zero staggering) for several values of $W$. We see that the parabolic approximation holds with an excellent accuracy for weaker disorder ($W=0.3, 0.5$ and 1.0), as expected (since the conductivity $\sigma$ in this case is high). For stronger disorder,  $W=2.0$ and 3.0, when the conductivity is not so large, deviations from parabolicity become clearly observable (although remain relatively small). The values of $\sigma$ and $\kappa$ extracted from the parabolic fits are presented in Table~\ref{tab:MF} as $\sigma_1$ and $\kappa_1$, respectively. The values of $x_\nu$ obtained from these fits are in good agreement with those found directly from the DOS scaling.

\begin{figure*}
	\centering
	\includegraphics[width=1.0\textwidth]{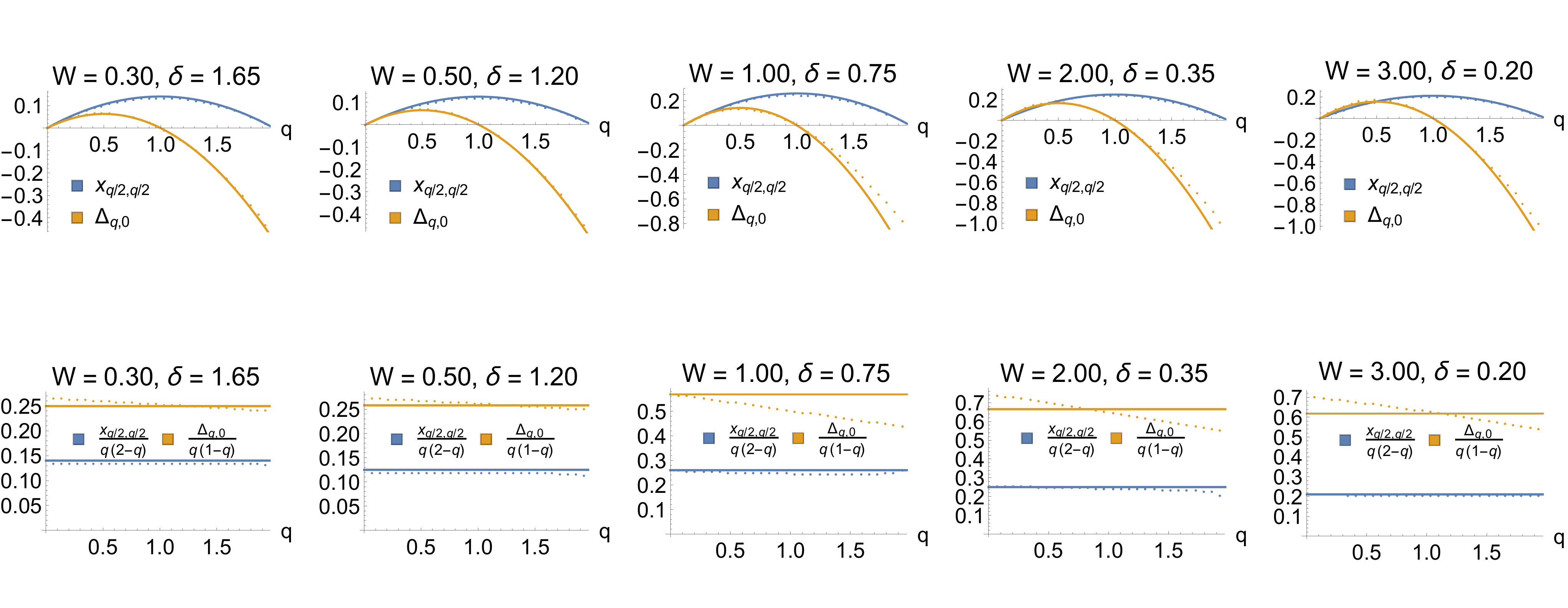}
	\caption{{\it Upper panels:} Multifractal exponents $\Delta_q$ (yellow dots) and $x_{q/2,q/2}$ (blue dots) as functions of $q$ at the MIT critical points (staggering $\delta \approx \delta_c$) for disorder strengths $W=0.3, 0.5, 1.0, 2.0, 3.0$.  The solid curves are the parabolic approximations~\eqref{eq:Delta-qs},~\eqref{eq:x-q2-q2s}.
	{\it Lower panels:} Same data presented as $\Delta_q/q(1-q)$  and $x_{q/2,q/2}/q(2-q)$, respectively. In this representation, the parabolic approximations become horizontal straight lines.}
	\label{fig:mf-critical}
\end{figure*}

Figure~\ref{fig:mf-critical} displays  the multifractal spectra $\Delta_q$ and $x_{q/2,q/2}$ at critical points of the MIT transition, $\delta \approx \delta_c(W)$. An apparent non-universality of the multifractal spectra at criticality is clearly seen; it is analogous to the apparent non-universality of other characteristics of the critical point ($\xi_M/M$, conductivity); reasons for it are discussed in the main text. We also observe a clear non-parabolicity of $\Delta_q$, which becomes particularly strong at critical points with stronger disorder ($W=1.0$, 2.0, 3.0), for which $\sigma$ is smaller. All these observations are consistent with the expected evolution of the spectra towards the ultimate infinite-randomness fixed point of the MIT, see the flow diagram in Fig.~\ref{fig:RG-flow} of the main text.

By means of the multifractal analysis we also determine $\alpha_0 = dx_q /dq |_{q=0}$, which is used in evaluation of the right-hand side of Eq.~\ref{eq:exp-map} shown in Fig.~\ref{fig:transmf_aiii} and in Table~\ref{tab:critical-parameters}. We find that the relation~\eqref{eq:exp-map} holds within the numerical accuracy of our analysis, indicating invariance of the critical theory under the exponential conformal map between the cylinder and plane geometries, see main text.

\end{document}